\documentclass[aps,twocolumn,prd,tightenlines,nofootinbib,superscriptaddress,floatfix]{revtex4}

\usepackage[english]{babel}
\usepackage[utf8]{inputenc}

\usepackage{lineno,hyperref}

\usepackage{dcolumn}
\usepackage{placeins} 
\usepackage{ragged2e}
\usepackage{etoolbox}
\usepackage{bm}
\usepackage{lipsum}
\usepackage{amsmath}
\usepackage{graphicx,epsfig,wrapfig}
\usepackage{subcaption}
\usepackage[compatibility=false]{caption}
\usepackage{subcaption}

\DeclareCaptionJustification{justified}{\justifying}
\usepackage{amsthm,mathrsfs}
\usepackage{makeidx}
\usepackage{amsfonts}
\usepackage{amssymb}
\usepackage{amsmath}
\usepackage{mathtools}
\usepackage{fixmath}
\usepackage{amsbsy}
\usepackage[capitalize]{cleveref}
\usepackage{mathrsfs}
\usepackage{slashed}
\usepackage{bm}
\usepackage{multirow}
\usepackage{booktabs}
\usepackage{adjustbox}
\usepackage{comment}
\usepackage{dsfont}
\usepackage{float}

\usepackage{color}
\usepackage[dvipsnames]{xcolor}
\usepackage[normalem]{ulem}

\definecolor{myGreen}{rgb}{0.2,0.72,0.2}
\definecolor{myGold}{rgb}{0.83,0.69,0.22}
\definecolor{darkgreen}{RGB}{0,120,0}
\renewcommand\sout{\bgroup \color[rgb]{0.55,0.00,0.99} \ULdepth=-.5ex \ULset}

\newcommand{\TLT}[0]{\mathbb{T}_{\mathrm{LT}}}
\newcommand{\TNLT}[0]{\mathbb{T}_{\mathrm{NLT}}}

\usepackage{simplewick}

\usepackage{multirow}
\usepackage{braket}
\usepackage{soul}

\newcommand{\rDer}[1]{\overset{\rightarrow}{#1}\phantom{\,}}
\newcommand{\lDer}[1]{\overset{\leftarrow}{#1}\phantom{\,}}

\renewcommand{\[}{\begin{equation}}
\renewcommand{\]}{\end{equation}}

\usepackage{tikz}
\usetikzlibrary{patterns,decorations.pathmorphing,decorations.pathreplacing}
\usepackage[compat=1.1.0]{tikz-feynman}

\allowdisplaybreaks

\modulolinenumbers[5]

\begin{document}

\title{Collinear matching for leading power gluon transverse momentum distributions} 

\author{Alessio Carmelo Alvaro}
\thanks{Electronic address: \href{mailto:alessiocarmelo.alvaro01@universitadipavia.it}{alessiocarmelo.alvaro01@universitadipavia.it}}

\affiliation{Dipartimento di Fisica ``A. Volta", Universit\`a degli Studi di Pavia, I-27100 Pavia, Italy}
\affiliation{Istituto Nazionale di Fisica Nucleare, Sezione di Pavia, I-27100 Pavia, Italy}

\author{Nanako Kato}
\thanks{Electronic address: \href{mailto:nanako.kato@dsf.unica.it}{nanako.kato@dsf.unica.it}}

\affiliation{Dipartimento di Fisica, Universit\`a degli Studi di Cagliari, Cittadella Universitaria, I-09042 Monserrato (CA), Italy}
\affiliation{Istituto Nazionale di Fisica Nucleare, Sezione di Cagliari, Cittadella Universitaria, I-09042 Monserrato (CA), Italy}

\author{Barbara Pasquini}
\thanks{Electronic address: \href{mailto:barbara.pasquini@unipv.it}{barbara.pasquini@unipv.it}}

\affiliation{Dipartimento di Fisica ``A. Volta", Universit\`a degli Studi di Pavia, I-27100 Pavia, Italy}
\affiliation{Istituto Nazionale di Fisica Nucleare, Sezione di Pavia, I-27100 Pavia, Italy}

\author{Cristian Pisano}
\thanks{Electronic address: \href{mailto:cristian.pisano@unica.it}{cristian.pisano@unica.it}}

\affiliation{Dipartimento di Fisica, Universit\`a degli Studi di Cagliari, Cittadella Universitaria, I-09042 Monserrato (CA), Italy}
\affiliation{Istituto Nazionale di Fisica Nucleare, Sezione di Cagliari, Cittadella Universitaria, I-09042 Monserrato (CA), Italy}

\author{Simone Rodini}

\thanks{Electronic address: \href{mailto:simone.rodini@unipv.it}{simone.rodini@unipv.it}}
\affiliation{Dipartimento di Fisica ``A. Volta", Universit\`a degli Studi di Pavia, I-27100 Pavia, Italy}
\affiliation{Istituto Nazionale di Fisica Nucleare, Sezione di Pavia, I-27100 Pavia, Italy}

\date{\today}

\begin{abstract}
We compute the tree-level and one-loop matching relations for leading power gluon transverse momentum dependent parton distribution functions. At tree level, working within the spinor formalism, we focus on twist-2 and twist-3 contributions, deriving the series of mass corrections for both T-even and T-odd distributions within the considered twist accuracy.
At one-loop accuracy, we extend the parton-in-parton framework to include contributions beyond the leading term in the small-$b$ expansion. Applying this methodology to the gluon sector, we obtain for the first time the Wandzura-Wilczek approximation for the gluon worm-gear T distribution. 
Furthermore, we develop a method to include the twist-2 contribution to the mass corrections in one-loop results and provide a closed-form expression for the mass series suitable for numerical implementations.
\end{abstract}

\maketitle

\section{Introduction}
In the study of the internal structure of the nucleon and its relation to nucleon phenomenology, the transverse motion of hadron constituents plays a fundamental role. At the current level of knowledge, we take into account this motion through the transverse momentum dependent (TMD) parton distribution functions (PDFs), which extend the standard collinear PDFs to the full 3D momentum space.
TMD distributions (TMDs for brevity) exist for both quarks and gluons and appear in the expressions for the soft part of cross sections in lepton-hadron and hadron-hadron collisions. 
The quark sector is well-established from both the theoretical and phenomenological perspectives. Currently, several reliable extractions of the unpolarized quark TMD have been carried out~\cite{Bacchetta:2024qre,Bacchetta:2025ara,Moos:2025sal,Barry:2025glq}. Existing data have also enabled extractions of the helicity TMD~\cite{Yang:2024drd,Bacchetta:2024yzl}, and there have been attempts to infer the behavior of essentially all the leading twist quark TMDs (see Ch.~5 of Ref.~\cite{Boussarie:2023izj} and references therein).  
On the theoretical front, the framework has recently advanced to the study of next-to-leading power (NLP) distributions~\cite{Rodini:2022wki}, encoded in the quark-gluon-quark correlator, and their inclusion in factorization theorems~\cite{Rodini:2023plb}.

In contrast, the gluon sector is less developed. While a parametrization of the gluon-gluon correlator has been known for over two decades~\cite{Mulders:2000sh}, a formal factorization theorem for gluon-induced processes has yet to be proven. However, by adopting factorization as a working hypothesis, several studies have successfully explored observables sensitive to gluon contributions in both unpolarized and polarized nucleons~\cite{Boer:2010zf,Boer:2012bt,Lansberg:2017dzg,Boer:2011kf,Qiu:2011ai,Kato:2024vzt}. 
In particular, the so-called Mulders-Rodrigues function~\cite{Mulders:2000sh}, describing the distribution of linearly polarized gluons inside an unpolarized nucleon, has been widely investigated. It corresponds to an interference between
$+1$ and $-1$ gluon helicity states that would be suppressed without transverse momentum. It has been shown that
it modifies the unpolarized cross sections for the production of scalar and pseudoscalar particles in different
ways, depending on their parity~\cite{Boer:2012bt,Boer:2011kf,Kato:2024vzt}. It can also give rise to azimuthal asymmetries in heavy quark pair or dijet production in electron-proton collisions~\cite{Boer:2010zf}, and in photon pair~\cite{Qiu:2011ai} or quarkonium-photon production~\cite{Lansberg:2017dzg} in hadronic collisions.
Moreover, significant attention has been devoted to inferring information on the gluon Sivers function~\cite{Sivers:1989cc}, that is the azimuthal distribution of unpolarized gluons inside a transversely polarized nucleon, through both gluon-gluon fusion in hadronic collisions and photon-gluon fusion in lepton-hadron scattering. These processes are central to the physics programs at the LHC and the future EIC~\cite{LHCspin:2025lvj,AbdulKhalek:2021gbh,Boer:2015vso,Kishore:2019fzb,Echevarria:2026vca,Zheng:2018ssm}.

By definition, TMDs depend on the fraction $x$ of the proton longitudinal momentum carried by the active parton and on the transverse separation of the fields $b$, which is the Fourier conjugate of the transverse momentum of the parton $p_T$. 
As non-perturbative objects, their functional form cannot be derived from first principles. However, in the small-$b$ regime, TMDs can be related to collinear PDFs through so-called matching relations.  
Under this framework, the functional form of TMDs is expressed as a Mellin convolution of appropriate collinear PDFs, supplemented by a nonperturbative part that needs to be extracted from experimental data.
Therefore, nearly all phenomenological studies on TMDs are grounded on these matching relations.
To date, these relations have been calculated to three-loop order for unpolarized~\cite{Ebert:2020yqt,Luo:2020epw} and helicity~\cite{Zhu:2025gts} distributions, as well as for quark transversity~\cite{Zhu:2025brn} and the gluon Mulders-Rodrigues function~\cite{Zhu:2025ixc}. For the remaining quark distributions, matching relations are known at one-loop~\cite{Rein:2022odl}, while the corresponding results in the gluon sector have not yet been analyzed.

Recently, many efforts have been devoted to the systematic inclusion of beyond leading power (LP) contributions in TMD processes. This involves separating higher-power contributions and systematically incorporating all contributions of a specific type~\cite{Vladimirov:2023aot,Piloneta:2025jjb}. 
Among these are the hadron mass corrections to factorization theorems. In their standard formulation, factorization theorems for TMD processes are typically derived in the limit of massless hadrons. 
To date, mass corrections have only been included at tree level in matching relations~\cite{Moos:2020wvd,Rodini:2023mnh} and, to the best of our knowledge, there are no proposals for extending these corrections to higher orders in perturbation theory.
Their inclusion could provide a refinement in the theoretical predictions and lead to a better analysis of existing and forthcoming experimental data.

Before detailing our framework, we clarify the different power expansions involved in our analysis. Our focus is on leading-power (LP) gluon TMDs. 
To establish their connection with collinear PDFs, we perform an Operator Product Expansion (OPE) of the LP TMD correlator around $b=0$. 
At tree level, this reduces to the Taylor expansion of the correlator. 
At one-loop level, our perturbative calculation generates a phase factor $e^{i(bp_T)}$, which acts as the generating function for an expansion in powers of $(bp_T)^n$. 
Upon Fourier transformation to position space, this maps directly onto the small-$b$ expansion. 
Each term in the OPE is expressed as a linear combination of collinear operators with definite geometric twist. 
Since retaining all twist components is impractical and contributions of twist-4 and higher are expected to be phenomenologically negligible for current applications, we truncate our analysis to selected twist components. 
Specifically, at tree level we retain the Leading Twist (LT) and Next-to-Leading Twist (NLT) contributions (which correspond to twist-2 and twist-3 operators, respectively), whereas at one-loop order we restrict our calculation to LT (twist-2) contributions. 
Within the specified twist accuracy, the procedure systematically generates hadron-mass corrections as an infinite series in $(M^2 b^2)^n$, where $M$ is the hadron mass.

This work is devoted to analyzing the full $b$ series of the gluon-gluon correlator up to one-loop accuracy. This analysis allows us to obtain the matching relations at tree-level and one-loop accuracies for the LP gluon TMDs and to include the hadron mass corrections beyond the tree-level approximation. In Sec.~\ref{sec:definitions}, we establish the fundamental definitions and conventions for the transverse and collinear distributions used throughout this study. 
In Sec.~\ref{sec:tree-level} we focus on the tree-level computation of the matching relations. By adopting the twist-decomposition technique within the spinor formalism, we derive the full mass dependence for both T-even and T-odd gluon TMDs up to next-to-leading twist PDFs. 
In Sec.~\ref{sec:one-loop}, we evaluate the one-loop matching relations by extending the standard parton-in-parton framework to incorporate higher-twist operators. This allows us to extract the one-loop matching coefficients, including the Wandzura-Wilczek approximation for the worm-gear T distribution. We also show that, within our framework, the inclusion of mass corrections at one-loop is  achieved straightforwardly. 
Finally, we summarize our findings and present our conclusions in Sec.~\ref{sec:conclusions}. The paper is complemented by three appendices. In Apps.~\ref{app:tree-level_example} and~\ref{app:one-loop_example} we provide detailed examples of tree-level and one-loop computations, respectively, while in  App.~\ref{app:sum_mass_series} we show the derivation of a closed form for the mass series.

\section{Definitions}\label{sec:definitions}
We start with a few definitions that we will use throughout the paper. We decompose a generic Lorentz four-vector $v^\mu$ as
\begin{align}
    v^\mu =\,& v^+ \bar{n}^\mu + v^- n^\mu + v_T^\mu \nonumber \\
    =\,& v^+ \bar{n}^\mu + v^- n^\mu - (vL) R^\mu - (vR) L^\mu\, ,
\end{align}
where $(ab)=a_\mu b^\mu$. The basis four-vectors $\{ \bar{n}, n, R, L\}$
are light-like and normalized such that $(n\bar{n}) = 1$, $(LR)=-1$, while all other scalar products vanish. A specific implementation of this basis is
\begin{align}
    n^\mu =& \frac{1}{\sqrt{2}} (1,0,0,-1) \, , \quad \, \, \bar{n}^\mu = \frac{1}{\sqrt{2}} (1,0,0,1) \, , \\
    R^\mu =& -\frac{1}{\sqrt{2}} (0,1,i,0) \, , \quad L^\mu = -\frac{1}{\sqrt{2}} (0,1,-i,0) \, .
\end{align}
The space spanned by the four-vectors $L, R$ is denoted as the transverse space, and four-vectors that have components only in this space are marked with a subscript $T$. Throughout this work, boldface notation denotes Euclidean scalar products in the transverse space, while standard font implies a Minkowski metric.
We denote the hadron momentum and spin four-vectors by $P$ and $S$, respectively, and the parton momentum by $p$ (such that $p^+=xP^+$ is the light-cone momentum of the parton). We have
\begin{align}
\label{eq:hadron-momentum}
    P^\mu =& P^+ \bar{n}^\mu + \frac{M^2}{2P^+} n^\mu \, ,\\
\label{eq:hadron-spin}
    S^\mu =& S_L \frac{P^+}{M} \bar{n}^\mu - S_L\frac{M}{P^+}n^\mu + S_T^\mu \, , \\
\label{eq:parton-momentum}
    p^\mu =& x P^+ \bar{n}^\mu - \frac{\boldsymbol{p}_{\boldsymbol{T}}^2}{2xP^+}n^\mu + p_T^\mu \, ,
\end{align}
where $M$ is the hadron mass and $S_L$ is the light-cone helicity of the hadron.
These four-vectors fulfill the relations $(PS)=0$, $P^2=M^2$, $S^2=-1$ and $p^2=0$.
We also define the metric and Levi-Civita tensors in the transverse plane as
\begin{align}
\label{eq:transverseMTD}
    g_T^{\mu\nu} =& g^{\mu\nu} - n^\mu \bar{n}^\nu - \bar{n}^\mu n^\nu = -R^\mu L^\nu - L^\mu R^\nu \, , \\
\label{eq:transverse-LeviCivita}
    \epsilon_T^{\mu\nu} =& \epsilon^{\alpha\beta\mu\nu} \bar{n}_\alpha n_\beta = i\left( R^\mu L^\nu - L^\mu R^\nu \right) \, ,
\end{align}
with the convention $\epsilon_T^{12}= \epsilon^{\alpha\beta 12} \bar{n}_\alpha n_\beta=+1$.

\subsection{Definition of transverse distributions}
In the literature, two distinct (LP) gluon-gluon TMD correlators are defined. They differ in the specific gauge link structure determined by the process in which the gluon TMDs appear~\cite{Bomhof:2006dp,Bomhof:2007xt,Buffing:2013kca}.  These  are typically referred to as the Weizsäcker-Williams (WW) and  dipole distributions~\cite{Kharzeev:2003wz,Boussarie:2023izj}.
Specifically, WW correlators are characterized by two gauge links or Wilson lines pointing in the same direction (i.e., either both future-pointing $[+,+]$ or both past-pointing $[-,-]$), whereas dipole correlators involve gauge links pointing in opposite directions ($[+,-]$ or $[-,+]$).  
We will focus on the correlator characterized by a WW-type gauge link configuration, which is relevant for a wide range of processes in both hadron-hadron and lepton-hadron collisions, such as Higgs production $gg\to H$~\cite{Echevarria:2015uaa} and the gluon-photon fusion with dijet production~\cite{delCastillo:2020omr,Echevarria:2026vca}. 
For simplicity we work in the adjoint representation, though our results are independent of the specific color representation.
The correlator reads
\begin{align}
    & G^{\mu\nu} (x,b) =\frac{1}{xP^+} \int \frac{dz}{2\pi}e^{-izxP^+}  \nonumber\\
    &\times \left\langle P, S \left| F^{\mu+}_a \left( zn + b \right) \,  \mathcal{W}_{\mp}^{ab} (z,b,\infty) \,  F^{\nu+}_b \left( 0 \right) \right| P, S \right\rangle \, ,
\label{eq:GG-correlator}\end{align}
where  $ b^\mu =(0,\boldsymbol{b_T},0)$ is the transverse separation of the partonic fields, $\mathcal{W}_{\mp}$ is the gauge link given by
\begin{align}
   & \mathcal{W}_{\mp}^{ab} (z,b,\infty) = \left[ zn + b, \mp\infty n + b \right]^{aa'} \nonumber\\
    &\times\left[ \mp\infty n + b, \mp\infty n \right]^{a'b'} \left[ \mp\infty n, 0 \right]^{b'b}\, ,
\end{align}
where the straight Wilson line connecting two spacetime points $v$ and $w$ is defined as
\begin{equation}
    [v,w]=\mathcal{P} \exp\left[-ig\int_0^1 dt \, (w^\mu - v^\mu) \, A^\mu(wt + (1-t)v) \right]\, .\label{eq:WL}
\end{equation}
In Eq.~\eqref{eq:WL}, the gluon field is implicitly contracted with the color generators $t^a$ in the adjoint representation. 
The sign $-$ ($+$) in the definition of the correlator is related to hadron-hadron (lepton-hadron) scattering processes.

For spin $1/2$ hadrons, the correlator in Eq.~\eqref{eq:GG-correlator} is parametrized by eight independent LP gluon TMDs. There exist several conventions in the literature~\cite{Boer:2016xqr, Boussarie:2023izj, Mulders:2000sh, Lorce:2013pza,Echevarria:2015uaa} and 
we adopt the decomposition presented in Ref.~\cite{Boussarie:2023izj}:
\begin{align}
\label{eq:unpolarized-correlator}
    2 G^{\mu\nu}_{U} (x,b) =& -g_T^{\mu\nu} \left( f_1^g + iM b_\alpha S_\beta \epsilon_T^{\alpha\beta} f_{1T}^{\perp g} \right)\, , \\
\label{eq:circularly-polarized-correlator}
    2 G^{\mu\nu}_{L} (x,b) =& -i\epsilon_T^{\mu\nu} \left( S_L g_{1L}^g + iM(bS)g_{1T}^g \right) \, ,\\
    2 G^{\mu\nu}_{T} (x,b) =& - \left( \frac{g_T^{\mu\nu}}{2} - \frac{b^\mu b^\nu}{b^2} \right) \frac{b^2M^2}{2} h_1^{\perp g} \nonumber \\
    &- S_L\frac{M^2}{4} b_{\alpha}\epsilon_T^{\alpha \{ \mu}b^{\nu\} } h_{1L}^{\perp g} \nonumber \\
    & + \frac{i}{12}M^3 (bS) b_{\alpha}\epsilon_T^{\alpha \{ \mu}b^{\nu\} } h_{1T}^{\perp g} \nonumber \\
    \label{eq:linearly-polarized-correlator}
    & + \frac{i}{4} M \left( b_{\alpha}\epsilon_T^{\alpha \{ \mu}S^{\nu\} } + S_{\alpha}\epsilon_T^{\alpha \{ \mu}b^{\nu\} } \right) h_{1T}^g\, ,
\end{align}
where $a^{\{ \mu} b^{\nu \}} = a^\mu b^\nu + a^\nu b^\mu$ and the subscripts $U$, $L$, $T$ refer to unpolarized, circularly polarized and linearly polarized gluons, respectively.

All the LP TMDs are real, dimensionless functions of $x$ and $b^2$.
They  are distinguished by their behavior under time-reversal (T) transformations.
Four of them (the  unpolarized $f_1^g$,  helicity $g_{1L}^g$, worm-gear T $g_{1T}^g$ and Mulders-Rodrigues $h_1^{\perp, \, g}$ distributions) are T-even, while the other four (the Sivers $f_{1T}^{\perp g}$, pseudo worm-gear L $h_{1L}^{\perp, \, g}$, transversity $h_{1T}^g$ and pretzelosity $h_{1T}^{\perp, \, g}$ distributions) are T-odd and therefore change sign depending on the direction of the gauge link. 
\subsection{Definition of collinear distributions}
In our computation, different LT and NLT collinear PDFs for spin $1/2$ hadrons are involved. 
We adopt the definitions of Refs.~\cite{Scimemi:2019gge,Rein:2022odl}, which are summarized here for clarity.
The LT gluon distributions are defined as
\begin{align}
    &\left\langle P,S \left| F^{\mu+} \left( zn \right) \left[ zn , 0 \right] F^{\nu+} \left( 0 \right) \right| P,S \right\rangle \nonumber \\
    &=- P_+^2\int_{-1}^1 dx e^{ixzP^+} \frac{x}{2} \left( g^{\mu\nu}_T f_g(x) + i S_L \epsilon_T^{\mu\nu} \Delta f_g(x) \right) \, .\nonumber\\
    &
\label{eq:LP-Gluon-PDF}
\end{align}
Here, $f_g$ ($\Delta f_g$) corresponds to the distribution of unpolarized (circularly polarized) gluons inside unpolarized (longitudinally polarized) hadrons. 
$f_g$ is an odd function of $x$, while $\Delta f_g$ is even. 
The LT quark distributions are
\begin{widetext}
\begin{align}
\label{eq:quark-unpolarized-PDF}
    \left\langle P,S \left| \bar{q} \left( zn  \right) \left[ zn, 0 \right] \gamma^+ q \left( 0  \right) \right| P,S \right\rangle =& 2P^+ \int_{-1}^1 dx e^{ixzP^+} f_1(x) \, , \\
\label{eq:quark-helicity-PDF}
    \left\langle P,S \left| \bar{q} \left( zn  \right) \left[ zn, 0 \right] \gamma^+ \gamma^5 q \left( 0  \right) \right| P,S \right\rangle =& 2S_L P^+ \int_{-1}^1 dx e^{ixzP^+} g_1(x) \, , \\
\label{eq:quark-transversity-PDF}
    \left\langle P,S \left| \bar{q} \left( zn  \right) \left[ zn, 0 \right] i\sigma^{\alpha+}\gamma^5 q \left( 0  \right) \right| P,S \right\rangle =& 2 S_T^\alpha P^+ \int_{-1}^1 dx e^{ixzP^+} h_1(x) \, .
\end{align}
\end{widetext}
The helicity distribution $g_1$ is an even function of $x$, while the unpolarized $f_1$ and the transversity $h_1$ are odd. 
Conventionally, PDFs are presented only for positive values of $x$, 
with the negative $x$ region mapped onto antiquark distributions at positive $x$.
In this work, we will keep the full $[-1,1]$ range for PDFs, to have a more uniform presentation between the LT and NLT cases.

For our calculations, we use only a specific subset of the NLT collinear PDFs defined in Refs.~\cite{Scimemi:2019gge,Rein:2022odl}. 
In the extraction of the NLT contributions (see App.~\ref{app:tree-level_example}), the only operation performed on the color matrices is their commutation. This operation results in a three-gluon operator proportional to the antisymmetric structure constant $f_{abc}$. Therefore, we do not require the $F_i^-$ PDFs, which are defined in terms of the symmetric $d_{abc}$  structure constant. 

Furthermore, the gluon equations of motion (EOMs) produce
only the  $\bar{q}\gamma^+ F^{\mu+} q$ combination among the three different quark-gluon-quark operators. This fact restricts the set of NLT quark PDFs onto which the gluon TMDs can match.
For brevity, we use the notation $f(x_{1,2,3})=f(x_1,x_2,x_3)$ and denote the integration measure as
    $\int [dx] = \int_{-1}^1 dx_1 \int_{-1}^1 dx_2 \int_{-1}^1 dx_3 \, \delta(x_1+x_2+x_3) \, $.
Due to the delta function in the integration measure, the NLT collinear distributions are in principle functions of only two independent momentum fractions. 
However, in the following we will keep explicitly the three momentum variables,
as this representation makes the underlying symmetry properties of the distributions more transparent.
Omitting the straight Wilson lines connecting the fields, we define the relevant correlators as follows:
\begin{widetext}
\begin{align}
\label{eq:NLP-gluon-PDF}
    \left\langle P,S \left| ig f^{abc} F_a^{\mu+}(z_1n) F_b^{\nu+}(z_2n) F_c^{\rho+}(z_3n)  \right| P,S \right\rangle =& M\left( P^+ \right)^3 \int [dx] e^{-i(x_1z_1+x_2z_2+x_3z_3)P^+} \sum_{i=2,4,6} t_i^{\mu\nu\rho} F^+_i(x_{1,2,3}) \, ,\\
\label{eq:NLP-quark-PDF}
    \left\langle P,S \left| g \bar{q}(z_1n) F^{\mu+}_a(z_2n)T^a \gamma^+ q(z_3n)  \right| P,S \right\rangle =& 2\epsilon_T^{\mu\nu} S_\nu M \left( P^+ \right)^2 \int [dx] e^{-i(x_1z_1+x_2z_2+x_3z_3)P^+} T(x_{1,2,3}) \, ,
\end{align}
\end{widetext}
where the tensor structures $t_i$ read 
\begin{align}
    t_2^{\mu\nu\rho} = & S_\alpha \left( \epsilon_T^{\mu\alpha} g_T^{\nu\rho} + \epsilon_T^{\nu\alpha} g_T^{\rho\mu} + \epsilon_T^{\rho\alpha} g_T^{\mu\nu} \right) \, , \\
    t_4^{\mu\nu\rho} = & S_\alpha \left( 2\epsilon_T^{\mu\alpha} g_T^{\nu\rho} - \epsilon_T^{\nu\alpha} g_T^{\rho\mu} -\epsilon_T^{\rho\alpha} g_T^{\mu\nu} \right) \, , \\
    t_6^{\mu\nu\rho} = & S_\alpha \left( \epsilon_T^{\mu\alpha} g_T^{\nu\rho} - \epsilon_T^{\rho \alpha} g_T^{\mu\nu} \right) \, . 
\end{align}
The NLT gluon PDFs are written in terms of two independent functions $G_+$ and $Y_+$ as follows
\begin{align}
    F_2^+(x_{1,2,3}) &= -\frac{1}{4} G_+(x_{1,2,3}) \, , \\
    F_4^+(x_{1,2,3}) &= -\frac{1}{2} Y_+(x_{1,2,3}) \, ,\\
    F_6^+(x_{1,2,3}) &= \frac{1}{2} \left( Y_+(x_{1,3,2}) - Y_+(x_{2,1,3}) \right) \, .
\end{align}
The NLT distributions satisfy the following symmetry relations
\begin{align}
\label{eq:prop1}
    T(x_{1,2,3}) = & \, T(-x_{3,2,1}) \, , \\
    \begin{split}
    G_+(x_{1,2,3}) = & \, G_+(-x_{3,2,1}) = - G_+(x_{2,1,3}) \\
    = & -G_+(x_{1,3,2}) \, , 
    \end{split} \\
    Y_+(x_{1,2,3}) = & \, Y_+(-x_{3,2,1}) = - Y_+(x_{3,2,1}) \, , \\
\label{eq:prop4}
    Y_+(x_{1,2,3}) \, + & \, Y_+(x_{2,3,1}) + Y_+(x_{3,1,2}) = 0 \, ,
\end{align}
where $f(-x_{3,2,1})=f(-x_3,-x_2,-x_1)$.

\section{Tree-level matching relations}
\label{sec:tree-level}
At tree-level, quantum fields may be treated as classical, and the OPE procedure simplifies to a Taylor expansion of the TMD correlator $G^{\mu\nu} (x,b)$ around the point $b=0$:
\begin{equation}
\label{eq:OPE-tree-level}
    G^{\mu\nu} (x,b) = \sum_{n=0}^\infty \frac{1}{n!} b_{\mu_1} \dots b_{\mu_n} \, \left( \partial^{\mu_1}_T \dots \partial^{\mu_n}_T \, G^{\mu\nu}(x,b) \right)|_{b=0} \, ,
\end{equation}
where the $n$-th term of the series 
corresponds to a combination of collinear operators with geometrical twist $t$, ranging in $2\le t \le n+2$.  
To isolate specific geometrical twist components from each term, we employ the twist-decomposition technique within the spinor formalism, following the approach established for quark TMDs in Refs.~\cite{Moos:2020wvd, Rodini:2023mnh}. This methodology allows us to derive the matching series with full mass dependence for each gluon TMD, up to terms proportional to NLT collinear distributions.

This section is  devoted to the tree-level computation. 
We start in Sec.~\ref{ss:spinor-formalism} with a compendium of definitions in the spinor formalism necessary for  our derivation. 
In Sec.~\ref{ss:twist-decomposition}, we describe the general framework  and, in Sec.~\ref{ss:tree-level_results}, we collect and discuss the results. Supplementary details are given in App.~\ref{app:tree-level_example},  where we show a step-by-step example of the computation, and in App.~\ref{app:sum_mass_series}, where we present a closed-form expression for the sum of the mass series.

\subsection{Spinor formalism}
\label{ss:spinor-formalism}
The spinor formalism\footnote{We use the conventions of Ref.~\cite{Moos:2020wvd}. For a comprehensive review of the spinor formalism and its applications in quantum field theory, see Refs.~\cite{Dreiner:2008tw,Sohnius:1985qm}.} is based on the local isomorphism between the Lorentz group $SO(3,1)$ and the group of complex unimodular matrices $SL(2,\mathbb{C})$. 
By virtue of this isomorphism, each four-vector $x^\mu$ is mapped to a Hermitian matrix as $x_{\alpha\dot{\alpha}} = x_\mu \sigma^\mu_{\alpha\dot{\alpha}}$, with $\sigma^\mu = (\mathds{1}, \sigma^1, \sigma^2, \sigma^3)$ and $\sigma^i$ being the Pauli matrices. 
Dotted indices belong to the conjugate representation $\left( u_\alpha \right)^*=\bar{u}_{\dot{\alpha}}$.
Under this convention, the scalar product of two vectors is given by $2x_\mu y^\mu = x_{\alpha\dot{\alpha}} y^{\dot{\alpha}\alpha}$.
Spinor indices are raised and lowered using the Levi-Civita tensor $\epsilon_{\alpha\beta}$
 $(\epsilon_{\dot{\alpha}\dot{\beta}})$, with the convention $\epsilon^{12}=-\epsilon^{\dot{1}\dot{2}}=1$.
The scalar product of two spinors is defined as $(uv)=-\epsilon_{\alpha\beta}u^\alpha v^\beta = -(vu)$, $(\bar{u}\bar{v})=-\epsilon_{\dot{\alpha}\dot{\beta}}\bar{u}^{\dot{\alpha}}\bar{v}^{\dot{\beta}}=-(\bar{v}\bar{u})$.
Defining a basis of spinors $\left\{\lambda, \mu, \bar{\lambda}, \bar{\mu}\right\}$, normalized according to $(\mu\lambda)(\bar{\lambda}\bar{\mu})=2$, a generic four-vector is decomposed as
\begin{equation}
    x_{\alpha\dot{\alpha}} = x^- \lambda_\alpha \bar{\lambda}_{\dot{\alpha}} + x^+ \mu_\alpha \bar{\mu}_{\dot{\alpha}} - x_T \mu_\alpha \bar{\lambda}_{\dot{\alpha}} - \bar{x}_T \lambda_\alpha \bar{\mu}_{\dot{\alpha}} \, .
\end{equation}
For brevity, we will use the notation $v_{\lambda\bar{\lambda}}=v_{\alpha\dot{\alpha}}\lambda^\alpha\bar{\lambda}_{\dot{\alpha}}$ (and similarly for other spinor combinations).
The gluon field strength tensor $F_a^{\mu\nu}$ is represented in this formalism as
\begin{equation}
    F_{\alpha\dot{\alpha}\beta\dot{\beta}}^a = 2(f^a_{\alpha\beta}\epsilon_{\dot{\alpha}\dot{\beta}}-\epsilon_{\alpha\beta}\bar{f}^a_{\dot{\alpha}\dot{\beta}})\, ,
\end{equation}
where $f_{\alpha\beta}$ is a symmetric spinor (with $\bar{f}=f^\dagger$) and $a$ is a color index. 
For our purposes, it is sufficient to consider only the good components $F^{\mu+}$, given by
\begin{equation}
    F_{\alpha\dot{\alpha}\lambda\bar{\lambda}}^a = \lambda_{\alpha}\bar{\lambda}_{\dot{\alpha}} \left( \frac{f^a_{\mu \lambda}}{(\mu \lambda)} + \frac{\bar{f}^a_{\bar{\mu}\bar{\lambda}}}{(\bar{\lambda\bar{\mu}})}  \right) - \mu_{\alpha}\bar{\lambda}_{\dot{\alpha}} \frac{f^a_{\lambda \lambda}}{(\mu \lambda)} - \lambda_{\alpha}\bar{\mu}_{\dot{\alpha}} \frac{\bar{f}^a_{\bar{\lambda} \bar{\lambda}}}{(\bar{\lambda}\bar{\mu})} .
\end{equation}
Finally, the EOMs for the gluon field can be written in terms of the spinors   $f$ and $\bar{f}$ as
\begin{align}
\label{eq:EOM1}
    &\rDer{D}_{\lambda\bar{\lambda}} \bar{f}^a_{\bar{\lambda}\bar{\mu}} - \rDer{D}_{\lambda\bar{\mu}} \bar{f}^a_{\bar{\lambda}\bar{\lambda}} = g \left( \bar{\lambda}\bar{\mu} \right) \bar{q} T^a \gamma^+ q \, , \\
\label{eq:EOM2}
    & \rDer{D}_{\lambda\bar{\lambda}} f^a_{\mu\lambda} - \rDer{D}_{\mu\bar{\lambda}} f^a_{\lambda\lambda} = g\left( \mu\lambda \right) \bar{q} T^a \gamma^+ q \, ,
\end{align}
where $\rDer{D}_\mu^{ab} = \rDer{\partial}_\mu \delta^{ab} - igA_\mu^c t_c^{ab}$ is the covariant derivative in the adjoint representation acting to the right. 

\subsection{Twist decomposition in spinor formalism}
\label{ss:twist-decomposition}
We consider the correlator in Eq.~\eqref{eq:GG-correlator} in position space. 
The operator is compactified, setting a finite length $L$ to the light-cone Wilson lines:
\begin{align}
    \mathcal{G}^{\mu\nu}(z,b) &= F^{\mu+}_a (zn+b) \mathcal{W}^{ab}_\mp(z,b,\infty)  F^{\nu+}_b(0) \nonumber\\
    &= \lim_{L\to\mp\infty} F^{\mu+}_a (zn+b) \mathcal{W}^{ab}(z,b,L)  F^{\nu+}_b(0) \, .  
\end{align}
The operator is then Taylor-expanded around $b=0$,
\begin{align}
  &  \mathcal{G}^{\mu\nu}(z,b,L) \nonumber\\
   &\qquad\quad =\sum_{n=0}^\infty \frac{1}{n!} \left( F^{\mu+} [zn,Ln] \right)_a (b\lDer{D})^n_{ab} \left( [Ln,0] F^{\nu+} \right)_b \, .  
\end{align}
Finally, by expanding the field strength tensors around the spacetime point $Ln$, we obtain
\begin{align}
\label{eq:starting-point}
    \mathcal{G}^{\mu\nu}(z,b) &= \lim_{L\to \mp\infty} \sum_{n=0}^\infty \frac{1}{n!} \sum_{s,t=0}^\infty \frac{w_1^s w_2^t}{s!t!}\nonumber\\
    &\times\left[\left( F^{\mu+} \lDer{D}_+^s \right)_a (b\lDer{D})^n_{ab} \left( \rDer{D}_+^t F^{\nu+} \right)_b\right](Ln) \nonumber\\
    &= \lim_{L\to \mp\infty} \sum_{n=0}^\infty \frac{1}{n!} \sum_{s,t=0}^\infty \frac{w_1^s w_2^t}{s!t!} O^{\mu\nu}_{s,n,t}(Ln)\, ,
\end{align}
where $w_1=z-L$ and $w_2=-L$. Equation~\eqref{eq:starting-point} is the starting point of the matching computation. 
Specific gluon polarizations are isolated by contracting Eq.~\eqref{eq:starting-point} with the appropriate projection operators from the set $\{ -g_T^{\mu\nu} , -i\epsilon_T^{\mu\nu}, R^\mu R^\nu + L^\mu L^\nu \}$, which correspond to unpolarized, circularly polarized, and linearly polarized gluons, respectively.

The next step consists in extracting the geometrical twist component from each operator $O^{\mu\nu}_{s,n,t}$ in Eq.~\eqref{eq:starting-point}.
In the spinor formalism, the twist decomposition is significantly simplified: due to the antisymmetry of the  scalar product of spinors, the decomposition is equivalent to the systematic (anti)symmetrization of spinor indices. In particular, the LP representation is obtained by totally symmetrizing all pairs of indices, while the NLP representation is 
derived by antisymmetrizing exactly one pair of indices while symmetrizing the remainder, and so forth for higher-order terms.
These operations are realized  by applying a differential projection operator $\mathbb{T}$, i.e., 
\begin{equation}
    O^{\alpha\dot{\alpha}\beta\dot{\beta}}_{s,n,t} = \sum_{k=0}^n \mathbb{T}_{\mathrm{N}^k\mathrm{LT}} O^{\alpha\dot{\alpha}\beta\dot{\beta}}_{s,n,t} \, .  
\end{equation}
In this work we compute the series up to $k=1$. The $k\ge2$ terms are related to NNLT and  higher-order collinear distributions which have yet to be defined.
The projectors $\mathbb{T}_{\mathrm{N}^k\mathrm{LT}}$ for $k=0, 1$, as  derived in Ref.~\cite{Moos:2020wvd}, are given by
\begin{align}
    &\TLT = \left( \partial^\mu_\lambda \right)^m \left( \partial^{\bar{\mu}}_{\bar{\lambda}} \right)^{\bar{m}} \frac{l!}{m!(l+m)!} \nonumber\\
        \label{eq:TLToperator}
   &\times \frac{\bar{l}!}{\bar{m}!(\bar{l}+\bar{m})!} \left( \partial^\lambda_\mu \right)^m \left( \partial^{\bar{\lambda}}_{\bar{\mu}} \right)^{\bar{m}}, \\
    &\TNLT^{(\bar{\mu}\bar{\lambda})} =\left( \partial^\mu_\lambda \right)^m \left( \partial^{\bar{\mu}}_{\bar{\lambda}} \right)^{\bar{m}-1} \frac{l!}{m!(l+m)!} \frac{(\bar{l}-1)!(\bar{l}+\bar{m}-1)}{(\bar{m}-1)!(\bar{l}+\bar{m})!}  \nonumber\\
    \label{eq:TNLToperator-barred}
    &\times \left[ \left( \partial^{\bar{\mu}}_{\bar{\mu}} \right)\left( \partial^{\bar{\lambda}}_{\bar{\lambda}} \right) + \left( \partial^{\bar{\mu}}_{\bar{\mu}} \right) - \left( \partial^{\bar{\mu}}_{\bar{\lambda}} \right)\left( \partial^{\bar{\lambda}}_{\bar{\mu}} \right) \right] \left( \partial^\lambda_\mu \right)^m \left( \partial^{\bar{\lambda}}_{\bar{\mu}} \right)^{\bar{m}-1} \, , 
  \\
    \label{eq:TNLToperator}
    &\TNLT^{(\mu\lambda)}  =\TNLT^{(\bar{\mu}\bar{\lambda})}(a\leftrightarrow \bar{a}) \, ,  
\end{align}
where $l$ ($m$) is the number of the occurrences of $\lambda$ ($\mu$) (and similarly $\bar{l}$ ($\bar{m}$) for their barred versions), $\TLT= \mathbb{T}_{\mathrm{N}^0\mathrm{LT}}$, $\TNLT=\TNLT^{(\bar{\mu}\bar{\lambda})} + \TNLT^{(\mu\lambda)}$ and 
\begin{equation}
    \partial_\lambda^\mu \equiv \mu \frac{\partial}{\partial \lambda} \, .
\end{equation}
The $\left( \partial^\lambda_\mu \right)^m \left( \partial^{\bar{\lambda}}_{\bar{\mu}} \right)^{\bar{m}}$ derivatives in Eq.~\eqref{eq:TLToperator} replace each 
occurrence of $\mu$ ($\bar{\mu}$) with $\lambda$ ($\bar{\lambda}$), symmetrizing the tensor. Then the action of the $\left( \partial^\mu_\lambda \right)^m \left( \partial^{\bar{\mu}}_{\bar{\lambda}} \right)^{\bar{m}}$ derivatives restores the original number of $\mu$'s and $\bar{\mu}$'s.
Similarly, in Eq.~\eqref{eq:TNLToperator-barred}, the operator in square brackets antisymmetrizes a pair of dotted indices, while the rest of the expression behaves as $\TLT$. In our calculation, the second set of derivatives $\partial_\lambda^\mu$ (and barred version) is applied after evaluating the forward matrix element.

We now project  Eq.~\eqref{eq:starting-point} into the spinor formalism:
\begin{widetext}
\begin{align}
    \mathcal{G}^{\alpha\dot{\alpha}\beta\dot{\beta}} =& \lim_{L\to\mp\infty} \sum_{n=0}^\infty \frac{(-1)^n}{n!} \sum_{s,t=0}^\infty \frac{z_1^s z_2^t}{s!t!2^{s+t+n}} \left( F^{\alpha\dot{\alpha}\lambda\bar{\lambda}} \lDer{D}_{\lambda\bar{\lambda}}^s \right)_a \left( b_T \lDer{D}_{\lambda\bar{\mu}} + \bar{b}_T \lDer{D}_{\mu\bar{\lambda}} \right)^n_{ab} \left( \rDer{D}^t_{\lambda\bar{\lambda}} F^{\beta\dot{\beta}\lambda\bar{\lambda}} \right)_b \nonumber\\
    =& \lim_{L\to\mp\infty} \sum_{n=0}^\infty \frac{(-1)^n}{n!} \sum_{s,t=0}^\infty \frac{z_1^s z_2^t}{s!t!2^{s+t+n}} O^{\alpha\dot{\alpha}\beta\dot{\beta}}_{s,n,t} \, .
\end{align}
For specific polarizations, we obtain
\begin{align}
    &\mathcal{G}_U = -g^{\mu\nu}_T \mathcal{G}_{\mu\nu} = \lim_{L\to\mp\infty} \sum_{n=0}^\infty \sum_{s,t=0}^\infty \frac{z_1^s z_2^t(-1)^n}{n!s!t!2^{s+t+n}} \left( O^{\bar{f}f}_{s,n,t} + O^{f\bar{f}}_{s,n,t} \right)\, , \\
    &\mathcal{G}_L = -i\epsilon_T^{\mu\nu} \mathcal{G}_{\mu\nu}= \lim_{L\to\mp\infty} \sum_{n=0}^\infty \sum_{s,t=0}^\infty \frac{z_1^s z_2^t(-1)^n}{n!s!t!2^{s+t+n}} \left( O^{\bar{f}f}_{s,n,t} - O^{f\bar{f}}_{s,n,t} \right) \, ,\\
    &\mathcal{G}_T = (R^\mu R^\nu + L^\mu L^\nu) \mathcal{G}_{\mu\nu} = \lim_{L\to\mp\infty} \sum_{n=0}^\infty \sum_{s,t=0}^\infty \frac{z_1^s z_2^t(-1)^n}{n!s!t!2^{s+t+n}} \left( O^{ff}_{s,n,t} + O^{\bar{f}\bar{f}}_{s,n,t} \right)\, ,
\end{align}
where $O^{ff}=\lambda_\alpha \bar{\mu}_{\dot{\alpha}} \lambda_\beta \bar{\mu}_{\dot{\beta}} O^{\alpha\dot{\alpha}\beta\dot{\beta}}$ (and similar for other contractions). 
We now apply the $\TLT$ and $\TNLT$ projectors to these correlators, evaluate the matrix elements, and Fourier transform the results to the $x$-space, as shown in detail in App.~\ref{app:tree-level_example}.

\subsection{Results}
\label{ss:tree-level_results}
The matching relations for the gluon TMDs onto the LT and NLT collinear PDFs read
\begin{align}
\label{eq:unpol}
    f_1^g(x,b) =& f_g(x) + \sum_{k=1}^\infty \frac{1}{k!(k-1)!} \left( \frac{x^2M^2b^2}{4} \right)^k \int_0^1 du \int dy \, \delta(x-uy) \left( \frac{\bar{u}}{u} \right)^{k-1} \, f_g(y) \, , \\
    f_{1T}^{\perp g} (x,b) =& \mp 2\pi \Bigg\{ \frac{\left(2F^+_2 + F^+_4\right)(-x,0,x)}{x} + \sum_{k=1}^{\infty} \frac{1}{(k-1)!(k+1)!} \left( \frac{x^2M^2b^2}{4}\right)^k  \nonumber  \\
    & \qquad\quad \times \int_0^1 du \int dy \, \delta(x-uy) \left( \frac{\bar{u}}{u} \right)^k \frac{1+(k-1)u+u^2}{\bar{u}}  \frac{\left(2F^+_2 + F^+_4\right)(-y,0,y)}{y} \Bigg\} \, ,
\label{eq:sivers}
\\
    g_{1L}^g (x,b) =& \Delta f_g(x) + \sum_{k=1}^{\infty} \frac{1}{k!(k-1)!} \left( \frac{x^2M^2b^2}{4} \right)^k \int_0^1 du \int dy \,  \delta(x-uy) \left( \frac{\bar{u}}{u} \right)^{k-1} \nonumber \\
    & \times \Bigg[ \left(1-2\bar{u} \, \phantom{}_2F_1(1,1,k+1;\bar{u}) \right) \Delta f_g(y)  + 2\frac{\bar{u}}{u} \left(1-\bar{u} \, \phantom{}_2F_1(1,1,k+2;\bar{u}) \right) \frac{\mathcal{F}(y)+ \mathcal{T}(y)}{y^2} \Bigg] \, ,
\label{eq:helicity} \\
    g_{1T}^{g} (x,b) =& -x \int_0^1 du \int dy \, \delta(x-uy) \Bigg\{ \Delta f_g(y) + \frac{\mathcal{F}(y) + \mathcal{T}(y)}{y^2} \nonumber \\
    & + \sum_{k=1}^\infty \frac{1}{k!k!} \left( \frac{x^2M^2b^2}{4} \right)^k \left( \frac{\bar{u}}{u} \right)^k \Bigg[ u \, \phantom{}_2F_1\left(1,1,k+1;\bar{u}\right) \, \Delta f_g(y) \nonumber \\
    & + \left( 1 + \frac{k}{(k+1)(k+2)} \frac{\bar{u}^2}{u} \phantom{}_2 F_1\left( 1,1,k+3;\bar{u} \right) \right) \frac{\mathcal{F}(y)+ \mathcal{T}(y)}{y^2} \Bigg] \Bigg\} \, ,
\label{eq:wormgearT} \\
    h_{1T}^g (x,b) =& \mp 2\pi \Bigg\{ \frac{\left( 2F^+_2 - 2F^+_4 \right)(-x,0,x)}{x} + \sum_{k=1}^\infty \frac{1}{k!(k-1)!}  \left( \frac{x^2M^2b^2}{4} \right)^k \nonumber \\
    & \qquad \quad \times \int_0^1 du \int dy \, \delta(x-uy) \left( \frac{\bar{u}}{u} \right)^{k} \left( \frac{u}{\bar{u}} + \frac{\bar{u}}{k+1} \right) \frac{\left( 2F^+_2 - 2F^+_4 \right) (-y,0,y)}{y} \Bigg\} \, ,
\label{eq:transversity}
 \\
    h_{1L}^{\perp g} (x,b) =& \mp 2\pi x \sum_{k=0}^\infty \frac{1}{k!k!} \left( \frac{x^2M^2b^2}{4} \right)^k \int_0^1 du \int dy \, \delta(x-uy) \left( \frac{\bar{u}}{u} \right)^k  \frac{u+k}{k+1} \frac{\left( 2F^+_2 - 2F^+_4 \right) (-y,0,y)}{y} \, ,
\label{eq:pseudo-wormgear}
\end{align}
where $\bar{u}=1-u$, $\phantom{}_2F_1(a,b,c;z)$ is the hypergeometric function and
\begin{equation}
\label{eq:tw3_convolution}
    \mathcal{F} (y) = \int \frac{[dy]}{y_2} \delta(y+y_1) \left[ \left( 2F^+_2 + F^+_4 \right) (y_{1,2,3}) + \left( 1 + \frac{y}{y_2} \right) F^+_6 (y_{1,2,3}) \right] \, , \qquad \mathcal{T} (y) = \int [dy] \, \delta(y+y_2) \, T(y_{1,2,3}) \, .
\end{equation}
\end{widetext}
Equations~\eqref{eq:unpol}-\eqref{eq:pseudo-wormgear} are expressed as Mellin convolutions and can be rewritten for numerical implementation as 
\begin{equation}
\begin{split}
    [G\otimes F](x) &= \int_0^1 du \int dy \, \delta (x-uy) \, G(u) F(y) \\
    &= \begin{cases}
        \int_x^1 \frac{dy}{y} G\left( \frac{x}{y} \right) F(y) \, , & x>0\, ,  \\
        \int_{-x}^1 \frac{dy}{y} G\left( \frac{-x}{y} \right) F(-y) \, , & x<0\, .
    \end{cases}
\end{split}
\end{equation}
All the series can be summed back to Bessel functions (see App.~\ref{app:sum_mass_series}). 
For those distributions matching onto collinear quark distributions, a sum over all the active flavors is understood.
In T-odd distributions the upper (lower) sign refers to hadron-hadron (lepton-hadron) scattering processes.

To the best of our knowledge, this is the first  systematic analysis of the complete $b$-series for the gluon TMD operator.
We correctly reproduce the well known leading terms  for the unpolarized~\eqref{eq:unpol} and helicity~\eqref{eq:helicity} TMDs. 
Furthermore, the leading terms for the worm-gear T~\eqref{eq:wormgearT}, Sivers~\eqref{eq:sivers} and transversity~\eqref{eq:transversity} TMDs which can be extracted from the matching of the operator $F^{\mu+}D^\alpha F^{\nu+}$ as reported in Ref.~\cite{Rein:2022odl} are in full agreement with our results. The remaining results, i.e.\ the leading term in the matching relation for the pseudo worm-gear L and the sets of mass series up to twist-3 accuracy, are presented here for the first time.

We note that all the mass series in Eqs.~\eqref{eq:unpol}-\eqref{eq:pseudo-wormgear} coincide exactly  with the mass series of the corresponding quark distribution.
This correspondence is immediately evident for the unpolarized~\eqref{eq:unpol} and the Sivers~\eqref{eq:sivers} distributions when compared,  respectively, with Eqs.~(4.1) and~(4.2) in Ref.~\cite{Moos:2020wvd}.
For the helicity and the worm-gear T distributions we need recasting the Lerch transcendent function in terms of the hypergeometric function. 
Using Eq.~(25.14.3) in Ref.~\cite{NIST:DLMF}, we have
\begin{align}
    \Psi_n(u) =& \Phi\left( \frac{\bar{u}}{\bar{u}-1}, 1, n \right) = \frac{1}{n} \phantom{}_2F_1\left( 1,n,n+1,\frac{\bar{u}}{\bar{u}-1} \right) \nonumber \\
    =& \frac{u}{n} \phantom{}_2 F_1\left( 1, 1, n+1, \bar{u} \right)\, ,
\end{align}
where, in the last step, we used~\eqref{eq:transformation-property-2F1}.
After the replacement of the Lerch function, Eqs.~\eqref{eq:helicity},~\eqref{eq:wormgearT} correspond exactly  to the quark results in  Eqs.~(4.3) and~(4.4) in Ref.~\cite{Moos:2020wvd}.
Finally, since the distributions for linearly polarized gluons match onto NLT collinear PDFs, we compare the mass series proportional to the LT PDF $h_1$ in Eqs.~(4.5) and~(4.7) of Ref.~\cite{Moos:2020wvd} with our results in Eqs.~\eqref{eq:transversity},~\eqref{eq:pseudo-wormgear}. It remains an open question whether this exact correspondence between the mass series of quark and gluon distributions is only due to an accidental correspondence in the operator structure or is the manifestation of a more profound underlying symmetry.

An interesting feature of Eqs.~\eqref{eq:unpol}-\eqref{eq:pseudo-wormgear} is the fact that all terms proportional to NLT collinear PDFs are accompanied by a  $1/x$ factor. 
Using the properties of collinear PDFs in Eqs.~\eqref{eq:prop1}-\eqref{eq:prop4}, one can show that for T-odd distributions, these terms take  the indeterminate form of the type $0/0$ in the limit $x\to 0$. A recent extraction of NLT collinear distributions~\cite{Vladimirov:2025qrh} suggests that, at small $x$, both $2F_2^++F_4^+$ and $2F_2^+-2F_4^+$ scale as $x^\alpha$ with $\alpha\sim4$. As a result, no divergence should occur around $x=0$ for the T-odd distributions $f_{1T}^{\perp,g}$ and $h_{1T}^g$.
In contrast, the small-$x$ behavior of T-even distributions remains unconstrained by current theoretical frameworks.

Now we briefly comment on each distribution.

The unpolarized distribution in Eq.~\eqref{eq:unpol} matches onto the unpolarized gluon PDF $f_g$. It does not contain any NLT collinear PDF in its expansion, as expected, since all the NLT collinear PDFs in Eqs.~\eqref{eq:NLP-gluon-PDF} and~\eqref{eq:NLP-quark-PDF} depend on the spin of the hadron.
However, its mass series could potentially receive further contributions from Mellin convolutions with NNLT (and beyond) collinear PDFs. 

The helicity distribution in Eq.~\eqref{eq:helicity} reduces to the circularly polarized gluon PDF $\Delta f_g$ in the  $b=0$ limit. Unlike the unpolarized TMD, its mass series receives contributions from both quark and gluon NLT collinear PDFs. However, even the first term in the mass series is incomplete and may  acquire additional corrections from NNLT collinear distributions.

We found no matching for the Mulders-Rodrigues distribution $h_{1}^{\perp g}$ at tree-level accuracy. There is no matching onto LT collinear PDFs since there is no counterpart of this distribution in the collinear sector. Furthermore, it does not match onto NLT distributions for the same reasons as the unpolarized case.
While we cannot a priori exclude a matching onto NNLT PDFs, the observed quark-gluon correspondence suggests that any such tree-level matching would likely mirror the mass series of the quark Boer-Mulders function.

The worm-gear T distribution in Eq.~\eqref{eq:wormgearT} matches onto both quark and gluon NLT PDFs already at its leading order. This is a unique feature, even if expected, resulting from the application of the gluon EOMs in Eqs.~\eqref{eq:EOM1},~\eqref{eq:EOM2}, which generate an explicit dependence on the quark fields in the operator.

The T-odd Sivers~\eqref{eq:sivers} and transversity~\eqref{eq:transversity} distributions share a similar structure. In these expressions, the upper (lower) sign refers to hadron-hadron (lepton-hadron) scattering processes. As already discussed above, the factor $1/x$ does not induce a divergence at $x=0$.

The pseudo worm-gear L distribution  in Eq.~\eqref{eq:pseudo-wormgear} matches onto the same gluon PDFs as the transversity TMD. Its leading term can be written as
\begin{equation}
    h_{1L}^{\perp g} (x,b) = x\left[u\otimes h_{1T}^{g}(x,b=0)\right](x) \, + \, O(\text{NNLT}) \, ,  
\end{equation}
in close analogy to 
the  Wandzura-Wilczek approximation of the $h_{1L}^\perp$ function in
the quark sector.  Since this distribution only begins to match at the $O(b^2)$ term in the Taylor expansion of the correlator, its leading term is incomplete and likely receives additional contributions from NNLT collinear PDFs. A similar scenario was observed for the quark pretzelosity in Ref.~\cite{Moos:2020wvd}.

Finally, we find no matching for the gluon pretzelosity $h_{1T}^{\perp g}$. Since the matching for this distribution starts at the $O(b^3)$ order in the Taylor expansion of the correlator (see App.~\ref{app:tree-level_example}), the leading term in its mass series is a combination of NNLT and NNNLT collinear distributions. 
Given the observed correspondence between quark and gluon mass series for the Mulders-Rodrigues function, we expect the first non-vanishing contribution to be identical to that of the quark pretzelosity.

\section{One-loop matching relations}\label{sec:one-loop}
The matching relation for the generic TMD operator $\Phi_{i}^\Lambda$, where $i=q,g$ denotes the parton flavor and $\Lambda=U,L,T$ its polarization, is  defined at any perturbative order by the following general expression
\begin{align}
    \Phi_{i}^\Lambda(x,b) =& \sum_{n=0}^{\infty} a_{\rm s}^n \sum_j \int_0^1 du\int dy \, \delta(x-uy) \nonumber \\ 
    \label{eq:generic-loop-matching-relation}
&\times\mathcal{C}_{ij}^{\Lambda/\Gamma,(n)}(u,b) \phi^\Gamma_{j}(y)\, ,
\end{align}
where $\phi_{j}^\Gamma$ is a collinear operator, $\mathcal{C}_{ij}^{\Lambda/\Gamma,(n)}$ is the matching coefficient connecting the TMD operator $\Phi$ with the collinear operator $\phi$ at order $a^n_{\rm s}$ and $a_{\rm s}=\alpha_{\rm s}/(4\pi)$. All the renormalization scales are omitted for brevity. 

Different techniques have been developed for the computation of these coefficients at higher-loop accuracy, including the parton-in-parton method~\cite{Collins:2011zzd,Bertone:2022awq,Bertone:2025vgy}, SCET calculations~\cite{Zhu:2025gts,Zhu:2025ixc} and background-field approach in position space~\cite{Scimemi:2019gge,Rein:2022odl}. 
Each of these approaches presents advantages and disadvantages. 
The position space approach allows one to find the matching onto higher-twists operators at the expense of significantly more complex computations; consequently, this technique has currently only been implemented for quark TMDs at one-loop order.
On the other hand, the parton-in-parton and SCET methods rely on standard momentum-space loop computation, making them better suited for higher-loop implementations. However, 
these methods have not yet been extended to TMDs that match onto operators beyond LT. 
Developing a framework to extend these techniques beyond LT is therefore essential to simplifying the study of matching relations; this is the  objective we aim to address. 
 In Sec.~\ref{ss:parton-in-parton}, we summarize the standard formulation of the parton-in-parton method. Then in Sec.~\ref{ss:extended-parton-in-parton}, we show how this approach is extended to include higher-twist operators. 
Finally, we collect the results obtained in this new framework in Sec.~\ref{ss:one-loop-results}. This section is further complemented by App.~\ref{app:one-loop_example} where we provide a detailed example of the computation.

\subsection{Standard parton-in-parton method}
\label{ss:parton-in-parton}
In the standard formulation of the parton-in-parton approach, the derivation of matching coefficients is cast as the computation of matrix elements of proper fields operators between partonic states, the so-called parton-in-parton TMDs. 
The parton-in-parton matrix elements are essentially obtained by replacing the hadronic states in the definition of the TMD operator in Eq.~\eqref{eq:GG-correlator} with massless, on-shell (un)polarized quark or gluon states. 
We stress that the validity of this approach relies on the factorization and universality of partonic distributions~\cite{Collins:2011zzd}. We work under this hypothesis although factorization has not yet been rigorously proven  for many processes involving gluon observables.

The gluon-in-gluon ($G_{gg}$) and gluon-in-quark ($G_{gq}$) distributions read
\begin{widetext}
\begin{align}
\label{eq:gluon-in-gluon}
    G_{gg}(x,b) =& -\frac{1}{2(N_c^2-1)} \int \frac{dz}{2\pi} \frac{e^{-ixzp_+}}{xp_+}  {\rm Tr} \left[ \left\langle p,s \left| F^{\mu+}_a(zn+b) \mathcal{W}^{ab}_{\mp}(z,b,\infty) \Lambda_{\mu\nu}  F^{\nu+}_b(0) \right| p,s \right\rangle_g \right] \, , \\
\label{eq:gluon-in-quark}
    G_{gq}(x,b) =& -\frac{1}{2N_c} \int \frac{dz}{2\pi} \frac{e^{-ixzp_+}}{xp_+}  {\rm Tr} \left[ \left\langle p,s \left| F^{\mu+}_a(zn+b) \mathcal{W}^{ab}_{\mp}(z,b,\infty) \Lambda_{\mu\nu}  F^{\nu+}_b(0) \right| p,s \right\rangle_q \right] \, .
\end{align}
\end{widetext}
The  subscripts $g$ and $q$ on the kets of the matrix elements denote the final partonic state of  momentum $p^\mu=xP_+\bar{n}^\mu$ and polarization $s$.  
The tensor  $\Lambda^{\mu\nu}\in\{ -g_T^{\mu\nu} , -i\epsilon_T^{\mu\nu}, R^\mu R^\nu + L^\mu L^\nu \}$ selects the initial gluon polarization and ${\rm Tr}$ refers to both Dirac and color traces. 
Since the fields are interacting, they can emit or absorb partons and change species before interacting with the final states.
Thus, these definitions can be used to compute, in principle at any order in $\alpha_{\rm s}$, the matching coefficients $\mathcal{C}$, by taking into account diagrams with additional radiation and using standard Feynman rules in momentum space.

From now on we will work in the light-cone gauge $n_\mu A^\mu=0$. This condition, however, does not completely fix  the gauge (see, e.g., Ref.~\cite{Scimemi:2019gge}) and it must be  supplemented by the boundary conditions $g^{\mu\nu}_T A_\nu(\mp\infty n)=0$, where the sign is chosen according to the direction of the Wilson line. 
With this choice, all Wilson lines in Eqs.~\eqref{eq:gluon-in-gluon} and~\eqref{eq:gluon-in-quark} reduce to the identity, so that the number of diagrams required for the calculation significantly decreases.
The choice of light-cone gauge leads to another simplification~\cite{Bertone:2022frx}: the field strength tensor reduces to $F^{\mu+}(x)=-(n\partial)A^\mu(x)$. When this operator acts on an external state with a momentum plus component  $p_+$, the derivative is replaced by a factor $ixp_+$ via the Fourier transform.
This is equivalent to the following replacement in Eqs.~\eqref{eq:gluon-in-gluon} and~\eqref{eq:gluon-in-quark}
\begin{equation}
    F^{\mu+}F^{\nu+} \to x^2 p_+^2 A^\mu A^\nu \, ,
\end{equation}
which simplifies loop calculations. In the light-cone gauge, therefore, the gluon parton-in-parton distributions become
\begin{widetext}
\begin{align}
    \label{eq:gluon-in-gluon-lg}
    G_{gg}(x,b) =& -\frac{xp_+}{2(N_c^2-1)} \int \frac{dz}{2\pi} e^{-ixzp_+} {\rm Tr} \left[ \left\langle p,s \left| A^\mu_a(zn+b) \Lambda_{\mu\nu} A^\nu_b(0) \right| p,s \right\rangle_g \right] \, , \\
\label{eq:gluon-in-quark-lg}
    G_{gq}(x,b) =& -\frac{xp_+}{2N_c} \int \frac{dz}{2\pi} e^{-ixzp_+} {\rm Tr} \left[ \left\langle p,s \left| A^\mu_a(zn+b) \Lambda_{\mu\nu} A^\nu_b(0) \right| p,s \right\rangle_q \right] \, .
\end{align}
\end{widetext}

\subsection{Extended parton-in-parton method}
\label{ss:extended-parton-in-parton}
In order to extend the parton-in-parton approach to higher-twist contributions, we start by considering partons with $\boldsymbol{p_T}\ne0$  in Eq.~\eqref{eq:parton-momentum}.

The extra mass dimension introduced by the non-vanishing transverse momentum can be compensated in the final result only by the transverse separation of the fields $b$, leading to terms of the form\footnote{In principle, terms of the form $(b^2p_T^2)^n$ could also appear. They effectively appear in intermediate steps of the computation; nonetheless, they cancel out in the final result. An example of this cancellation is provided in App.~\ref{app:one-loop_example}.} $(bp_T)^n$. These terms are related to higher-twist contributions and to show it, in the following, we explicitly recover the Taylor expansion in Eq.~\eqref{eq:OPE-tree-level}.

First, we need to define proper projectors $\Gamma^{\mu\nu} \in \left\{ d^{\mu\nu}(p) \, , -i\epsilon^{\mu\nu}(p) \, ,  t^{\mu\nu}(p) \right\}$ to isolate, respectively, unpolarized, circularly polarized and linearly polarized gluon final states when $p_T\ne0$.
Their expressions are
\begin{align}
    d^{\mu\nu}(p) =& -g^{\mu\nu} + \frac{p^\mu n^\nu + n^\mu p^\nu}{p_+} \nonumber \\
\label{eq:unpolarized-gluon-projector}
    =& -g^{\mu\nu}_T + \frac{p_T^\mu n^\nu + n^\mu p_T^\nu}{p_+} +2\frac{p_-}{p_+}n^\mu n^\nu \, , \\
\label{eq:polarized-gluon-projector}
    -i\epsilon^{\mu\nu}(p) =& -i\epsilon^{\mu\nu\alpha\beta}p_\alpha n_\beta \nonumber  \\
    =& -i\epsilon_T^{\mu\nu} -i\frac{\epsilon^{\mu\nu\alpha\beta}p_{T,\alpha} n_\beta}{p_+} \, , \\
    t^{\mu\nu}(p) =& R^\mu R^\nu + L^\mu L^\nu + \frac{\tilde{p}_T^\mu n^\nu + n^\mu \tilde{p}_T^\nu}{p_+} \nonumber \\
\label{eq:linearly-polarized-gluon-projector}
    &+ \frac{(pL)^2+(pR)^2}{p_+^2} n^\mu n^\nu \, .
\end{align}
The tensors $g_T^{\mu\nu}$ and $\epsilon_T^{\mu\nu}$ are defined in Eqs.~\eqref{eq:transverseMTD} and~\eqref{eq:transverse-LeviCivita} and $\tilde{p}_T^\mu \equiv -(pR)R^\mu - (pL)L^\mu$. Each projector is systematically separated into its leading, suppressed, and doubly suppressed components. Specifically,  the leading parts consist of the purely transverse tensors (i.e., $-g_T^{\mu\nu}$, $-i\epsilon_T^{\mu\nu}$, and $R^\mu R^\nu+L^\mu L^\nu$). These projectors are an extension of those used in Ref.~\cite{Bertone:2023jeh,Bertone:2025vgy} where the terms proportional to $p_T$ were neglected.

The derivation of these projectors is as follows. The final result of any $O(\alpha^n)$ calculation with gluon final states is proportional to the tensor $\epsilon^*_\mu \epsilon_\nu$.
In the light-cone gauge, the polarization vector $\epsilon_\mu$ associated with the gluon field $A_\mu$ is orthogonal to both the gluon momentum $p$ (by definition) and the vector $n$ (by the gauge condition). As a result, the tensor $\epsilon_\mu^*\epsilon_\nu$ lives in the Minkowski subspace orthogonal to both $n$ and $p$. 
The projector in Eq.~\eqref{eq:unpolarized-gluon-projector} corresponds to the metric tensor of this subspace and therefore selects unpolarized gluons. Similarly, Eq.~\eqref{eq:polarized-gluon-projector} corresponds to the Levi-Civita tensor within this subspace for circularly polarized gluons, while Eq.~\eqref{eq:linearly-polarized-gluon-projector}  extends the description of linear polarization to the same subspace.

Having defined  the projectors onto the gluon final states, we return to the matching computation. While Eq.~\eqref{eq:gluon-in-quark-lg} evaluates to zero at tree level, the result for the gluon-in-gluon case in Eq.~\eqref{eq:gluon-in-gluon-lg} follows by noting that
\begin{equation}
    A_a^\mu(x)| k,s \rangle = e^{-i(kx)} \epsilon_a^\mu(k,s)|0\rangle,
\end{equation}
and then  projecting the initial and final polarizations using $\Lambda_{\mu\nu}$ and $\Gamma^{\mu\nu}$. This leads to
\begin{equation}
\label{eq:tree-level-parton-in-parton}
    \mathcal{C}_{gg}^{U/U} = \mathcal{C}_{gg}^{L/L} = \mathcal{C}_{gg}^{T/T} = e^{i(bp_T)}\delta(1-u) \, .  
\end{equation}
We omit the superscript $(0)$ on $\mathcal{C}$ as it is clear we are referring to the tree-level coefficients.
Since the transverse momentum in position space corresponds to the derivative operator $p_T^\mu = -i\partial_T^\mu$, we can interpret the global phase in Eq.~\eqref{eq:tree-level-parton-in-parton} as the generating function for the Taylor series around $b=0$. Substituting this result into the tree-level term of Eq.~\eqref{eq:generic-loop-matching-relation} yields
\begin{align}
    G_{\Lambda}(x,b) =& \delta_{\Lambda\Gamma} \, \left( e^{i(bp_T)} G_{\Gamma}(x,b) \right)\nonumber \\
    =& \sum_{n=0}^\infty \frac{\delta_{\Lambda\Gamma}}{n!} b_{\mu_1} \dots b_{\mu_n} \left[ \partial_T^{\mu_1}\dots \partial_T^{\mu_n} G_{\Gamma}(x,b) \right]_{|b=0}\, ,\nonumber\\
 &   
\label{eq:tree-level-parton-in-parton-series}
\end{align}
that coincides with~\eqref{eq:OPE-tree-level}. In particular, it is worth noting that
\begin{align}
\label{eq:operator-FT}
   & \left[ \partial_T^{\mu_1}\dots \partial_T^{\mu_n} G_{\Gamma}(x,b) \right] \nonumber\\
    &\qquad\qquad= \int \frac{dz}{xP_+} e^{-ixzP_+} \left[ \partial_T^{\mu_1}\dots \partial_T^{\mu_n} G_{\Gamma}(z,b) \right]\, .
\end{align}
This implies that the decomposition in terms of PDFs is known for each operator of the series in Eq.~\eqref{eq:tree-level-parton-in-parton-series}, as already derived in Sec.~\ref{sec:tree-level}. 
In our one-loop analysis, we will truncate the expansion at LT, since matching onto NLT collinear distributions also requires the evaluation of diagrams  with three external partons. This is left for future work.

We have further verified that we can recover known results of position-space computations by applying this extended parton-in-parton approach, even at intermediate steps; for the sake of brevity, these explicit calculations are not reported here.
Specifically, we focused on the one-loop matching for T-even quark TMDs onto LT quark PDFs, evaluating diagrams (A) and (B) in Fig.~1 of~\cite{Rein:2022odl} using our framework in the Feynman gauge. For each diagram, we identified  a LT contribution, which can be expanded in  $O(b^0)$ and $O(b)$ terms, and an NLT contribution. We have explicitly checked that there is an exact correspondence term by term between the two approaches.

\subsection{One-loop results}
\label{ss:one-loop-results}
In this section, we present the results for the one-loop matching relations. Before discussing the explicit formulas, we clarify that our current one-loop calculation is restricted to diagrams with two external partons and does not incorporate contributions from diagrams with three external partons. The complete calculation, including three-parton contributions, is deferred to future work.
For brevity, we define
\begin{equation}
    L_b= \log\frac{\mu^2}{\mu_b^2} \, , \quad l_\zeta = \log\frac{\mu^2}{\zeta}\, ,
\end{equation}
where $\mu$ and $\zeta$ denote the dimensional regularization and rapidity evolution scales, respectively. Here $\mu_b$ is defined via $\boldsymbol{b}_{\boldsymbol{T}}^2\mu_b^2=4e^{-2\gamma_E}$ with $\gamma_E$ the Euler-Mascheroni constant. 
With these definitions, the general form of the matching coefficient at one-loop reads~\cite{Gutierrez-Reyes:2017glx,Bertone:2022frx,Bertone:2025vgy}
\begin{widetext}
\begin{align}
    \mathcal{C}_{ij}^{\Lambda/\Gamma}(u,b;\mu,\zeta) = -\delta_{\Lambda\Gamma}\mathcal{P}_{ij}^\Gamma(u) L_b + \mathcal{R}_{ij}^{\Lambda/\Gamma}(u) - \delta_{ij}\delta_{\Lambda\Gamma} \delta(1-u) C_i \left[ L_b^2 - 2\left( K_i + l_\zeta \right)L_b + \frac{\pi^2}{6} \right]\, ,
\end{align}
\end{widetext}
where $\mathcal{P}_{ij}^\Gamma$ are the LO evolution kernels, $\mathcal{R}_{ij}^{\Lambda/\Gamma}$ are the residual functions and
\begin{equation}
\begin{split}
    C_q=&C_F \, , \quad K_q=\frac{3}{2} \, , \\
    C_g=&C_A \, , \quad K_g=\frac{11}{6} - \frac{2T_R}{3C_A}n_f \, .
\end{split}
\end{equation}
The splitting functions at one-loop are  well-known results and are recovered here as a byproduct of our computation, providing a partial check on the correctness of our results. On the other hand, the residual functions within the extended parton-in-parton approach are obtained here for the first time. 
In a nutshell, keeping a non zero transverse momentum  $\boldsymbol{p_T}$ of the parton generates a global phase $e^{iu(bp_T)}$ in the matrix elements~\eqref{eq:gluon-in-gluon-lg} and~\eqref{eq:gluon-in-quark-lg} so that the extended matching coefficients are related to the standard ones by  $\mathcal{C}_{std}^{(1)}\to\mathcal{C}_{ext}^{(1)}=\mathcal{C}_{std}^{(1)} \, e^{iu(bp_T)}$, where $\mathcal{C}_{std}^{(1)}$ can be found or deduced from the results in Refs.~\cite{Bertone:2022awq,Bertone:2025vgy,Gutierrez-Reyes:2017glx,Rein:2022odl}. 
For more details on the computation, see App.~\ref{app:one-loop_example}.

Replacing the matching coefficients $\mathcal{C}_{ext}$ in the $O(a_{\rm s})$ term of the series in Eq.~\eqref{eq:generic-loop-matching-relation} leads to
\begin{widetext}
\begin{equation}
\label{eq:one-loop-series}
    \Phi_i^\Lambda (x,b) = a_{\rm s} \int_0^1 du \int dy \, \delta(x-uy) \sum\nolimits_j\mathcal{C}_{ij,std}^{\Lambda/\Gamma}(u,b) \sum_{n=0}^{\infty} \frac{u^n}{n!} b_{\mu_1}\dots b_{\mu_n} \left( \partial_T^{\mu_1}\dots \partial_T^{\mu_n} \phi_j^\Gamma(x,b) \right)_{|b=0}\, ,
\end{equation}
\end{widetext}
and, using Eq.~\eqref{eq:operator-FT} (and similar for quark operators), we recover the matching relations for T-even gluon distributions by replacing each operator in the series with its decomposition in terms of collinear distributions. These decompositions were derived in Sec.~\ref{sec:tree-level} for the gluon and in Ref.~\cite{Moos:2020wvd} for the quark. 
Note that Eq.~\eqref{eq:one-loop-series} suggests a systematic method for incorporating  hadron mass corrections into the TMD matching relations. We will investigate it in more detail at the end of this section.

At the end of the computation, we obtain the following results for the leading terms of the matching relations: 
\begin{widetext}
\begin{align}
\begin{split}
\label{eq:unpolarized-one-loop}
    f_1^g(x,b;\mu,\zeta) =& \left[ 1- a_{\rm s} C_A \left( L_b^2 -2K_g L_b -2L_b l_\zeta + \frac{\pi^2}{6} \right) \right] f_g(x) \\ 
    &- 2a_{\rm s} \int_0^1 du \int dy \, \delta(x-uy) \left[ 2C_AL_b \, p_{gg}(u) \, f_g(y) + C_F\left(L_b \, p_{gq}(u) - u \right)f_1(y)  \right] \, ,  
\end{split} \\
\label{eq:boer-mulders-one-loop}
    h_1^{\perp g}(x,b;\mu,\zeta) =& -4a_{\rm s} \int_0^1 du \int dy \, \delta(x-uy) \, \frac{\bar{u}}{u} \left( C_A f_g(y) + C_F f_1(y) \right) \, \\
\begin{split}
\label{eq:helicity-one-loop}
    g_{1L}^g(x,b;\mu,\zeta) =& \left[ 1- a_{\rm s} C_A \left( L_b^2 -2K_g L_b -2L_b l_\zeta + \frac{\pi^2}{6} \right) \right] \Delta f_g(x) \\
    & -2a_{\rm s} \int_0^1 du \int dy \, \delta(x-uy) \left[ 2C_A\left( L_b \, \Delta p_{gg}(u) + 2\bar{u} \right)\Delta f_g(y) + C_F\left( L_b \, \Delta p_{gq}(u) + 2\bar{u} \right)g_1(y) \right] \, ,
\end{split} \\
\begin{split}
\label{eq:worm-gear-T-one-loop}
    g_{1T}^g(x,b;\mu,\zeta) =& -\left[ 1- a_{\rm s} C_A \left( L_b^2 -2K_g L_b -2L_b l_\zeta + \frac{\pi^2}{6} \right) \right] x \, g_T^g(x) \\
    &+ 2a_{\rm s} \int_0^1 du \int dy \, \delta(x-uy) \, u \left[ 2C_A\left( L_b \, \Delta p_{gg}(u) + 2\bar{u} \right) y\, g_T^g(y) - C_F\left( L_b \, \Delta p_{gq}(u) + 2\bar{u} \right) y\, g_T(y) \right] \, ,
\end{split}
\end{align}
\end{widetext}
where 
\begin{align}
    g_T^g(x) =& \int_0^1 du \int dy \, \delta(x-uy) \, \Delta f_g(y) + O(\text{NLT})\, ,  \\
    g_T(x) =& \int_0^1 du \int dy \, \delta(x-uy) \, g_1(y) + O(\text{NLT}) \, .
\end{align}
The functions  $\Delta p_{gg}$ and $\Delta p_{gq}$ ($p_{gg}$ and $p_{gq}$) are the LO splitting functions for circularly polarized (unpolarized) gluons. Their expressions can be found in Refs.~\cite{Bertone:2022frx,Bertone:2023jeh,Gutierrez-Reyes:2017glx}.

The result for the worm-gear T $g_{1T}^g$ TMD is presented here for the first time. 
As noted at the beginning of this section, we report only the Wandzura-Wilczek approximation, as our current one-loop framework is restricted to two-parton diagrams.
In the quark case~\cite{Rein:2022odl}, it has been observed that the
matching of the worm-gear T and of the helicity TMDs onto the quark 
helicity PDF are (essentially) the same, up to an extra Mellin convolution. Equations~\eqref{eq:helicity-one-loop} and~\eqref{eq:worm-gear-T-one-loop} show a similar behavior in the gluon sector:
\begin{align}
\label{eq:relation-helicity-wormgearT}
    g_{1T}^g(x,b)_{|\text{gluon}} = -x\int_0^1 du \, g_{1L}^g(u,b)_{|\text{gluon}} \, .
\end{align}
In position space calculations, this arises from what appears to be a fortuitous cancellation between
the $O(b)$ part of the LT term and the NLT term of diagrams (A) and (B) in Fig.~1 of Ref.~\cite{Rein:2022odl}. 
In our approach instead this feature follows directly from the
phase factor $e^{iu(bp_T)}$ that emerges during the computation of 
the parton-in-parton matrix element. 
Since a corresponding phase is expected to appear in higher-loop 
calculations as well, we  speculate that this relation between 
helicity and worm-gear T TMDs should hold to all orders in 
perturbation theory within the Wandzura-Wilczek approximation.

We stress, however, that genuine higher-twist contributions, which are  neglected here, could potentially spoil this relation. Indeed, a relation such as Eq.~\eqref{eq:relation-helicity-wormgearT} no longer holds once twist-3 contributions are included, and this is already evident at tree-level accuracy: the leading term of the helicity distribution in Eq.~\eqref{eq:helicity} receives no twist-3 contribution, whereas the leading term in the worm-gear $T$ distribution in Eq.~\eqref{eq:wormgearT} does.

A similar argument can be extended to the quark sector, relating  helicity to  worm-gear T TMD and transversity to worm-gear L TMD.

We also report that we have obtained non-vanishing results for the matching coefficients $\mathcal{C}_{gg}^{T/T}$ and $\mathcal{C}_{gg}^{U/T}$. In particular, the former contains both a pole part 
(corresponding to the evolution kernel for the PDF of linearly polarized gluons in a transversely polarized hadron)
 and a finite part. 
Previous studies on generalized parton distributions (GPDs)~\cite{Bertone:2023jeh} and generalized transverse momentum dependent distributions (GTMDs)~\cite{Bertone:2025vgy} have already observed that the splitting function $\mathcal{P}_{gg}^T$ and the residual functions $\mathcal{R}_{gg}^{U/T}$ and $\mathcal{R}_{gg}^{T/T}$ are non-zero in the forward limit. Our results agree with these findings.
This fact seems to be in contradiction with the absence of a collinear distribution corresponding to linearly polarized gluons in a spin 1/2 hadron like the proton. However, this contradiction is  only apparent. 
As expressed in Eq.~\eqref{eq:generic-loop-matching-relation}, 
the matching relation is governed by the interplay between the matching coefficients  $\mathcal{C}$ and the collinear operator $\phi$. The dependence on the hadron is fully contained in the collinear operators, while 
the matching coefficients contain only  information  about the initial and final parton states.  Consequently, the coefficients are universal, i.e., valid for a generic hadron. In the specific case of the proton, $\phi_g^T$ vanishes, leading to zero matching relations between the gluon polarizations $U/T$ and $T/T$.

Finally, we return  to the hadron mass corrections beyond leading order. In Eq.~\eqref{eq:one-loop-series} we have found an expansion for the TMD correlator at one loop that includes all the powers of $b$. 
Since the operators  on the r.h.s.  are the Fourier transforms of the collinear operators in position space,  we can  replace each operator with its twist decomposition. By comparing this with the general decomposition of the gluon correlator, we can extract the mass series at one-loop and twist-2 accuracies. 
We show the final result for the mass series of the unpolarized TMD. Only the  $n=2k$ terms in Eq.~\eqref{eq:one-loop-series} contribute and substituting each operator with its decomposition yields the following one-loop mass series
\begin{widetext}
\begin{align}
    f_{1}^g(x,b) =& \int_0^1 du \int d\xi \, \delta(x-u\xi) \sum_{k=1}^\infty \frac{u^{2k}}{k!(k-1)!} \left( \frac{\xi^2M^2b^2}{4} \right)^k \nonumber \\
    &\times \int_0^1 dv\int dy \, \delta(\xi-vy) \left( \frac{\bar{v}}{v} \right)^{k-1} \left( \mathcal{C}_{gg}^{U/U}(u,b) f_g(y) + \mathcal{C}_{gq}^{U/U}(u,b)f_1(y) \right) \nonumber \\
\label{eq:one-loop-mass-series}
    =& \sum_{k=1}^{\infty} \left( \frac{x^2M^2b^2}{4} \right)^k \int_0^1 du\int_0^1 dv \int dy \, \frac{\delta(x-uvy)}{k!(k-1)!} \left( \frac{\bar{v}}{v} \right)^{k-1} \left( \mathcal{C}_{gg}^{U/U}(u,b) f_g(y) + \mathcal{C}_{gq}^{U/U}(u,b)f_1(y) \right) \, ,
\end{align}
\end{widetext}
where for the quark part we have used Eq.~(4.1) in Ref.~\cite{Moos:2020wvd} and $\bar{v}=1-v$.
Using Eq.~\eqref{eq:sum-unpolarized}, the infinite series can be expressed in terms of a Bessel function, providing a closed-form expression suitable for numerical implementation.

As a final remark, based on the structure of Eq.~\eqref{eq:one-loop-mass-series}, we deduce that at a generic order $O(\alpha^n)$ the mass series for a given TMD $F_i$ takes the form
\begin{align}
    &F_i(x,b) = \sum_{k=1}^\infty \frac{1}{k!k!} \left( \frac{x^2M^2b^2}{4} \right)^k \left(\prod_{l=1}^n \int_0^1 du_l\right) \int_0^1 dv \nonumber \\
    & \times\int_{-1}^1 dy \, \delta \left(x-vy\prod_{l=1}^nu_l \right) \, \sum\nolimits_j G_j(k,v) \mathcal{C}^{(n)}_{ij}(u_l,b) f_j(y) \, ,   
\end{align}
where $f_j$ is a LT PDF.
In this framework, each additional loop order adds
an extra Mellin convolution in $du_l$ to the Mellin convolution $dv$ of the mass series.

\renewcommand{\arraystretch}{1.3}

\begin{table*}[ht]
\centering
\begin{tabular}{c|c|c|c|c|c}
    Distribution & Twist of leading matching & Tw2 & Tw3 & Accuracy & Ref. \\
\hline
    $f_1^g$ & Tw2 & $f_g \, , f_1$ & - & N\textsuperscript{3}LO & \cite{Ebert:2020yqt,Luo:2020epw} \\
    $h_1^{\perp g}$ & Tw2 & $f_g \, , f_1$ & - & N\textsuperscript{3}LO & \cite{Zhu:2025ixc} \\
    $g_{1L}^g$ & Tw2 & $\Delta f_g \, , g_1$ & -  & N\textsuperscript{3}LO & \cite{Zhu:2025gts} \\
    $g_{1T}^g$ & Tw2-3 & $\Delta f_g \, , g_1$  & $\mathcal{F} \, , \mathcal{T}$   & NLO/LO  & This work \\
\hline 
    $f_{1T}^{\perp g}$ & Tw3 & - & $2F_2^++F_4^+$  & LO & This work \\
    $h_{1T}^g$ & Tw3 & - & $2F_2^+-2F_4^+$  & LO & This work \\ 
    $h_{1L}^{\perp g}$ & Tw3-4 & -  & $2F_2^+-2F_4^+$  & LO & This work \\
    $h_{1T}^{\perp g}$ & Tw4-5 & -  & -  & LO & This work \\
\end{tabular}
    \caption{Summary of  matching results for gluon TMDs, combining our new derivations with existing results from the literature. The upper (lower) part of the table corresponds to T-even (T-odd) distributions. Dashes denote cases where matching onto PDFs is absent at a given twist. 
    In the second-to-last column, the double entries 
    for the worm-gear T distribution
    indicate different accuracy level for the twist-2 and twist-3 components. Terms arising exclusively from the mass series are not listed. 
    In the last column, we list the references where individual results were computed.
    }
    \label{tab:summary}
\end{table*}

\section{Conclusions}
\label{sec:conclusions}

In this work, we analyzed the small-$b$ expansion of the gluon-gluon correlator (Eq.~\eqref{eq:GG-correlator}) at both  tree-level and  one-loop order, providing a comprehensive summary of the new and existing results in Tab.~\ref{tab:summary}. 
At tree level, we derived  the matching relations of LP TMDs onto LT and NLT collinear distributions (Eqs.~\eqref{eq:unpol}-\eqref{eq:pseudo-wormgear}), with our most remarkable new results being the matching relations for T-odd TMD distributions. Furthermore, at one-loop accuracy, by restricting our computation to diagrams with two external partons, we obtained the Wandzura-Wilczek approximation for the gluon worm-gear T distribution for the first time (Eqs.~\eqref{eq:unpolarized-one-loop}-\eqref{eq:worm-gear-T-one-loop}).

From a methodological perspective, the main result of this work is the extension of the parton-in-parton approach to include higher-twist operators. As shown in App.~\ref{app:one-loop_example}, the inclusion of the transverse momentum of the parton does not cause any significant increase in the complexity of the computation compared to the existing literature~\cite{Bertone:2022awq}, at least at one-loop.
We expect, therefore, that existing higher-loops calculation for TMDs matching onto LT parton distribution functions can be extended to include all T-even TMDs in both the quark and gluon sectors, paving the way for a more accurate analysis of higher-twist effects in TMD physics.
At present, this extension is restricted to the matching onto LT PDFs; however, we expect that NLT (and beyond) PDFs can be incorporated by computing diagrams with three or more external partons. Similarly, NLP TMDs could be included by employing a proper parton-in-parton definition.
Furthermore, with minor adjustments in the intermediate steps of the computation, this approach can be directly applied to GTMD matching relations.

On the phenomenological side, the matching relations and the Wandzura-Wilczek approximation for the worm-gear T distribution provide a robust theoretical framework that constrains the extraction of gluon TMDs, directly informing ongoing and future experimental efforts at the LHC and the EIC.
We have also shown that,  in the extended parton-in-parton framework, the inclusion of mass corrections at one loop follows straightforwardly. 
More specifically, the structure of Eq.~\eqref{eq:one-loop-series} reveals a significant procedural advantage: the mass series at one-loop accuracy can be systematically reconstructed by combining the tree-level twist decomposition with the matching coefficients derived from two-parton states.
In this work, the inclusion of mass corrections has been demonstrated at twist-2 accuracy. 
Twist accuracy can, in principle, be systematically improved by including diagrams with three or more external partons.
By providing a method to resum this infinite series into a closed form, alongside an ansatz for the higher-loop mass series, these results can be implemented in existing codes for TMD phenomenology, allowing for the first systematic inclusion of mass corrections in the analysis of processes involving TMDs or their extractions. \\

\section{Acknowledgments}
The work of N.K.\ and C.P.\ is supported by Fondazione di Sardegna through the project {\it Journey to the center of the proton}, No.\ F23C25000150007.
\appendix

\section{Example of tree-level NLT computation}
\label{app:tree-level_example}
In this appendix, we provide an example of tree-level computation. 
We have chosen the transversity distribution $h_{1T}^{g}$; due to its T-odd nature, this example illustrates all the essential steps required to reconstruct the full set of results presented in Sec.~\ref{ss:tree-level_results}.

The first step consists of applying $\TNLT$ to $O_{s,n,t}^{ff}$, which is conceptually straightforward but algebraically involved. Thus, we report only a sketch of the derivation  and the final result.
The antisymmetrization of the spinor indices produces operators of the form $F_TD^K_+F_+$ and $F_+ D_+^J D_T D_+^K F_+$, where $T$ denotes a transverse component. 
Using the gluon EOMs in Eqs.~\eqref{eq:EOM1} and~\eqref{eq:EOM2}, we replace the transverse component of the field with a transverse derivative.
Note that the application of the EOMs generates the quark contributions in the TMDs for circularly polarized gluons.
We then apply the following commutation relations
\begin{align}
\label{eq:comm-Dbarmu}
    \left[\lDer{D}_{\lambda\bar{\lambda}}^J, \lDer{D}_{\lambda\bar{\mu}}\right] &= 2ig (\bar{\lambda}\bar{\mu})\sum_{m=0}^{J-1} \lDer{D}_{\lambda\bar{\lambda}}^m f_{\lambda\lambda} \lDer{D}_{\lambda\bar{\lambda}}^{J-1-m} \, , \\
\label{eq:comm-Dmu}
    \left[\lDer{D}_{\lambda\bar{\lambda}}^J, \lDer{D}_{\mu\bar{\lambda}}\right] &= 2ig(\mu\lambda)\sum_{m=0}^{J-1} \lDer{D}_{\lambda\bar{\lambda}}^m \bar{f}_{\bar{\lambda}\bar{\lambda}} \lDer{D}_{\lambda\bar{\lambda}}^{J-1-m} \, .
\end{align}
After this step, all the bad components cancel out. The final result is
\begin{widetext}
\begin{align}
\label{eq:TNLT-Off}
    \TNLT^{(\mu\lambda)} O^{ff}_{s,n,t} =& -\sum_{k=1}^n \frac{(n-1)!(-1)^t}{(k-1)!(n-k)!} b_T^k \bar{b}_T^{n-k} \left( \partial^\mu_\lambda  \right)^{k-1} \left( \partial^{\bar{\mu}}_{\bar{\lambda}} \right)^{n-k}  \nonumber\\
    & \times \frac{(s+t+k)!}{(s+t+n)!} \frac{(s+t+n-k+3)!(s+t+n+3)}{(s+t+n+4)!}  \nonumber\\
    &\times \Bigg\{ 2ng(\mu\lambda) \left[\left(\bar{q}\gamma^+q)\lDer{D}_{\lambda\bar{\lambda}}^{s+t+n-1}\right)^a f^a_{\lambda\lambda} + \left(f_{\lambda\lambda} \lDer{D}_{\lambda\bar{\lambda}}^{s+t+n-1}\right)^a \left(\bar{q}\gamma^+ q\right)^a \right]   \nonumber\\
    & -2ig(\mu\lambda)(if^{abc}) \Bigg[ (s+t+n+4)n\sum_{m=0}^{s-2} + (s+t+n+4)\sum_{m=s-1}^{s+n-2}(s+n-1-m) \nonumber \\
    & + n\sum_{m=0}^{s+t+n-2} (s+t+n-m)\Bigg] \left( f_{\lambda\lambda} \lDer{D}_{\lambda\bar{\lambda}}^m \right)^a \bar{f}_{\bar{\lambda}\bar{\lambda}}^b \left( \lDer{D}_{\lambda\bar{\lambda}}^{s+t+n-2-m}f_{\lambda\lambda} \right)^c \Bigg\} \, .
\end{align}
\end{widetext}
The operator $\TNLT^{(\bar{\mu}\bar{\lambda})}O^{ff}_{s,n,t}$ vanishes when taking the matrix element and is therefore dropped. Next, we evaluate the matrix element, focusing only on the singular terms. Regular terms should be related to the only available T-even distribution, the Mulders-Rodrigues distribution $h_1^{\perp g}$, which, however, has zero matching at tree level, as  discussed in Sec.~\ref{ss:tree-level_results}.
The singular term reads
\begin{align}
    &  \braket{P,S|\TNLT^{(\mu\lambda)}O^{ff} |P,S}_{g}   \nonumber\\
    &=\lim_{L\to\mp\infty} \sum_{n=1}^{\infty} \sum_{s,t=0}^{\infty} \frac{(iw_1)^s(iw_2)^t(-1)^s}{s!t!2^{s+t+n+1}n} \sum_{k=1}^n \frac{b_T^k \bar{b}_T^{n-k} i^{n-2} }{(k-1)!(n-k)!}  \nonumber \\
    & \times \frac{(s+t+k)!}{(s+t+n)!} \frac{(s+t+n-k+3)!}{(s+t+n+2)!}  \nonumber \\
    & \times \left( \partial^\mu_\lambda  \right)^{k-1} \left( \partial^{\bar{\mu}}_{\bar{\lambda}} \right)^{n-k}  \frac{-iMS_{\lambda\bar{\mu}}}{2} P_{\lambda\bar{\lambda}}^{s+t+n+1}  \nonumber \\
    & \times \int [dy] \left( 2F^+_2 - 2F_4^+ \right)(y_{123}) \frac{y_1^s(-y_3)^t\left( y_1^n - (-y_3)^n \right)}{y_2^2}.
\end{align}
This term is related to the transversity distribution $h_{1T}^{g}$, the (pseudo)worm-gear L distribution $h_{1L}^{\perp g}$, and the pretzelosity distribution $h_{1T}^{\perp g}$. In particular, setting $n=2l$ and $\delta_{k,l+1}$, we obtain the matching relation for the worm-gear L distribution (all the other values of $k$ as a function of $l$ give vanishing results due to the action of derivatives).
For odd values of $n$, we obtain matching relations for the transversity (starting from $n=1$) and the pretzelosity (from $n=3$) distribution. However, in the latter case, there is no condition on $n$ and $k$ for which the action of the derivatives (see below) yields a non-vanishing result. This implies that the matching for the pretzelosity distribution starts with twist-4 PDFs.

Focusing on the transversity distribution and setting $n=2k-1$, we have
\begin{align}
    &\lim_{L\to\mp\infty} \sum_{k=1}^\infty \sum_{s,t=0}^\infty \frac{(-iw_1)^s(iw_2)^t}{s!t!2^{s+t+2k}(2k-1)} \frac{b_T^k \bar{b}_T^{k-1}i^{2k-3}}{(k-1)!(k-1)!}  \nonumber\\
    &\times \frac{(s+t+k)!(s+t+k+2)!}{(s+t+2k-1)!(s+t+2k+1)!} \left( \partial_\lambda^\mu \right)^{k-1} \left( \partial_{\bar{\lambda}}^{\bar{\mu}} \right)^{k-1}   \nonumber\\
    &\times \, P_{\lambda\bar{\lambda}}^{s+t+2k} \left( -\frac{MS_{\lambda\bar{\mu}}}{2} \right)  \nonumber\\
    &\times\int [dy] \frac{y_1^s(-y_3)^t\left( y_1^{2k-1} + y_3^{2k-1} \right)}{y_2^2} \left( 2F_2^+ - 2F_4^+ \right)(y_{1,2,3}) \, .
\end{align}
The derivatives are solved using the following identity
\begin{equation}
    \left( \partial^\mu_\lambda \right)^r \left( \partial^{\bar{\mu}}_{\bar{\lambda}} \right)^r P_{\lambda\bar{\lambda}}^{N+r} = \frac{r!(N+r)!}{N!} P_{\lambda\bar{\lambda}}^{N} P_{\mu\bar\mu}^r \, .
\end{equation}
which leads to
\begin{align}
\label{eq:step-transversity}
    &\lim_{L\to\mp\infty} \sum_{k=1}^\infty \frac{iMb_TS_T }{2k-1} \sum_{s,t=0}^\infty \left( \frac{M^2b^2}{4} \right)^{k-1}  \nonumber \\
    &\times \left( 1 - \frac{k-1}{s+t+2k+1} \right) \beta(s+t+k+1,k-1) \,  \nonumber \\
    &\times \frac{(-iP_+w_1)^s(iP_+w_2)^t}{s!t!(k-1)!(k-2)!} P_+^2  \nonumber \\
    &\times \int [dy] \frac{y_1^s(-y_3)^t\left( y_1^{2k-1} + y_3^{2k-1} \right)}{y_2^2} \left( 2F_2^+ - 2F_4^+ \right)(y_{1,2,3})\, ,
\end{align}
where
\begin{equation}
\label{eq:beta-function}
    \beta(l+1,m+1) = \frac{l!m!}{(l+m+1)!} = \int_0^1 du \, u^l \, \bar{u}^m \, ,
\end{equation}
with  $\bar{u}=1-u$.
Note that the series appears to be undefined at $k=1$. However, by expressing the factorials as Gamma functions and taking the limit $k\rightarrow 1^+$, we obtain ($N=s+t+k+1$)
\begin{equation}
    \lim_{k\to 1^+} \frac{\beta(N,k-1)(k-1)}{\Gamma(k-1)\Gamma(k)} = 0 \, , \quad \lim_{k\to1^+} \frac{\beta(N,k-1)}{\Gamma(k-1)}=1 \, .
\end{equation}
As a result, the  $k=1$ term of the series does not require any special treatment, as we can directly replace the beta function with its integral representation~\eqref{eq:beta-function}. 
On the other hand, terms with $k\ge2$ require further manipulation of the second factor in the second line of~\eqref{eq:step-transversity}. Since $1/(X+1) = \int_0^1 d\alpha \, \alpha^X$, we can write
\begin{equation} 
    (k-1)\int_0^1 d\alpha \int_0^1 du \, (\alpha u)^{s+t+k} \, \alpha^k \, \bar{u}^{k-2} \, ,
\end{equation}
which can be recast in a more convenient  form using the change of variable $\gamma=\alpha u$ 
\begin{equation}
    (k-1)\int_0^1 d\alpha \int_0^\alpha d\gamma \, \gamma^{s+t+k} \, \alpha \, (\alpha-\gamma)^{k-2} \, .
\end{equation}
The integral over $\alpha$ is then evaluated using the identity 
\begin{equation}
\label{eq:integral}
    \int_u^1 dx \, x^l \, (x-u)^n = \frac{\bar{u}^{n+1}}{n+1} \, \phantom{}_2F_1(1,-l,n+2;\bar{u}) \, .
\end{equation}
This result is derived at the end of the appendix.
For negative values of its second argument, the hypergeometric function reduces to a polynomial. In particular
\begin{equation}
    \phantom{}_2F_1(1,-1,k;\bar{u}) = 1 - \frac{\bar{u}}{k} \, .
\end{equation}
The second line of~\eqref{eq:step-transversity} can therefore be written as
\begin{equation}
    \int_0^1 du \, u^{s+t+k} \bar{u}^{k-1} \left( \frac{u}{\bar{u}} + \frac{\bar{u}}{k} \right) \, .
\end{equation}
At the end of this procedure (and restoring the $k=1$ term), we obtain
\begin{widetext}
\begin{equation}
\begin{split}
    \braket{P,S|\TNLT^{(\mu\lambda)}O^{ff} |P,S}_{g}^{sing} = &-\lim_{L\to\mp\infty} iMb_TS_T P_+^2 \int [dy] e^{-iP_+(w_1y_1+w_2y_3)} \frac{\left( 2F^+_2 - 2F_4^+ \right)(y_{1,2,3})}{y_2} \\
    &+ \lim_{L\to\mp\infty} \sum_{k=2}^\infty \frac{iMb_TS_TP_+^2}{(2k-1)(k-1)!(k-2)!} \left( \frac{M^2b^2}{4} \right)^{k-1}  \\
    &\times \int_0^1 du \int [dy] e^{-iuP_+(w_1y_1+w_2y_3)} u^k\bar{u}^{k-1} \left( \frac{u}{\bar{u}} + \frac{\bar{u}}{k} \right)  \\
    &\times \left( y_1^{2k-1} + y_3^{2k-1} \right) \frac{\left( 2F^+_2 - 2F_4^+ \right)(y_{1,2,3})}{y_2^2}.
\end{split}
\end{equation}
\end{widetext}
This expression is $L$-dependent: in fact, the exponent can be rewritten in terms of $z$ and $L$ as $w_1y_1 + w_2y_3 = zy_1 + Ly_2$\footnote{In T-even distributions this exponent is independent of $L$ and therefore the limit $L\rightarrow \pm \infty$ is trivial.}. 
The integral over the momentum fractions can be written as
\begin{align}
    &\int [dy] e^{-iuP_+(w_1y_1 + w_2y_3)} \nonumber  \\
    &=(-iuP_+) \int [dy] \, y_2 \, e^{-iuP_+zy_1} \int_{-L+z}^{L} d\tau \frac{e^{-iuP_+\tau y_2}}{2} \, .
\end{align}
The limit can now be safely taken, since\footnote{There is a typo in the sign of Eq.~(3.41) of~\cite{Moos:2020wvd}. However, the final results are correct and consistent with the literature.}
\begin{equation}
    \lim_{L\to\mp\infty} (-iuP_+) \int_{-L+z}^{L} d\tau \frac{e^{-iuP_+\tau y_2}}{2} = \pm i \pi \delta(y_2) \, .
\end{equation}
This step shows how the sign flip, characteristic of T-odd distributions, arises in our calculation. The integral over $y_2$ is evaluated with the help of the following relation
\begin{align}
    &\sum_{m=0}^{k-1} \frac{(2k-1)(2k-2m-2)!}{m!(2k-2m-1)!} \left( -y_1 y_3 \right)^m y_2^{2k-2m-2}  \nonumber\\
    &=-\frac{y_1^{2k-1} + y_3^{2k-1}}{y_2} \, .
\end{align}
Due to the delta function in $y_2$, only the term $m=k-1$ survives. 
The final step consists of the Fourier transform to the $x$-space. 
To obtain Eq.~\eqref{eq:transversity}, the computation needs to be completed with the analysis of $\TNLT^{(\bar{\mu}\bar{\lambda})}O^{\bar{f}\bar{f}}$.

As a final remark, we show how to solve the integral~\eqref{eq:integral}. 
This integral generates the hypergeometric functions found in the mass series of the helicity and worm-gear T TMDs.
Using the change of variables $x=u+\bar{u}v$, we have
\begin{equation}
     \bar{u}^{n+1} u^l \int_0^1 dv \,  v^{n+1-1} \bar{v}^{n+2-(n+1)-1} \left( 1 - \frac{\bar{u}}{\bar{u}-1}v \right)^{-(-l)},
\end{equation}
which can be related to the integral representation of the hypergeometric 
function:
\begin{equation}
\label{eq:integral-repres-2F1}
    \phantom{}_2F_1(a,b,c;z) = \int_0^1 dv  \frac{v^{b-1} \bar{v}^{c-b-1}}{\beta(b,c-b)} (1-zv)^{-a} \, .
\end{equation}
We obtain
\begin{equation}
    \frac{\bar{u}^{n+1}u^l}{n+1} \phantom{}_2F_1\left( -l,n+1,n+2; \frac{\bar{u}}{\bar{u}-1} \right)\, .
\end{equation}
Using the Pfaff transformation formula, Eq.~(15.8.1) in Ref.~\cite{NIST:DLMF}:
\begin{equation}
\label{eq:transformation-property-2F1}
    \phantom{}_2F_1(a,b,c,z) = (1-z)^{-b} \, \phantom{}_2F_1\left( b, c-a, c, \frac{z}{z-1} \right)   
\end{equation}
with $a=1$, $b=-l$, $c=n+2$ and $z=\bar{u}$,
we arrive at the expression in Eq.~\eqref{eq:integral}.

\section{Summation of the mass series}
\label{app:sum_mass_series}
The numerical implementation of the mass corrections in Eqs.~\eqref{eq:unpol}-\eqref{eq:pseudo-wormgear} requires a closed-form expression for the series. 
In this appendix we describe a general strategy to derive such expressions for all  TMDs. Note that this strategy can also be applied to the results in Refs.~\cite{Moos:2020wvd,Rodini:2023mnh}. 

As already noted in Ref.~\cite{Moos:2020wvd}, the mass series are suppressed by a double factorial and, since $b^2<0$, the expansion parameter is negative. 
This type of oscillating series resembles the Taylor expansion of the Bessel function of the first kind $J_n(\rho)$. 
Based on this observation, a closed form for  $f_1^g$, $f_{1T}^{\perp g}$, $h_{1T}^{g}$ and $h_{1L}^{\perp g}$ follows straightforwardly. For example, the unpolarized TMD~\eqref{eq:unpol} becomes
\begin{equation}
\label{eq:sum-unpolarized}
    f_1^g(x,b) = f_g(x) - \int_0^1 du \int dy \, \delta(x-uy) \, \frac{u}{\bar{u}} \rho \, J_1\left(2\rho \right)f_g(y) \, .
\end{equation}
where
\begin{equation}
    \rho=\frac{xM|b|}{2} \sqrt{\frac{\bar{u}}{u}} \, ,
\end{equation}
with $|b|=\sqrt{-b^2}$.
For the circularly polarized gluon distributions $g_{1L}^g$ and $g_{1T}^g$, the procedure is further complicated by the presence of the hypergeometric function. 
The steps in this case are as follows.
First, the hypergeometric function is recast into the form $\phantom{}_2F_1(a,b,b+1,z)$ using the Pfaff transformation~\eqref{eq:transformation-property-2F1}.
Second, we employ the integral representation of the hypergeometric function~\eqref{eq:integral-repres-2F1}. 
Third, 
we exchange the order of summation and integration and
perform the sum over the resulting Bessel series.
We illustrate these steps for the LP mass series of the helicity distribution~\eqref{eq:helicity}:
\begin{align}
    & \sum_{k=1}^{\infty} \frac{(-1)^k\rho^{2k}}{k!(k-1)!} u \, \phantom{}_2F_1 \left( 1,1,k+1,\bar{u} \right)  \nonumber\\
    & =\sum_{k=1}^{\infty} \frac{(-1)^k\rho^{2k}}{k!(k-1)!} \phantom{}_2F_1 \left( 1,k,k+1,\frac{\bar{u}}{\bar{u}-1} \right)  \nonumber \\
    & =\sum_{k=1}^{\infty} \frac{(-1)^k\rho^{2k}}{(k-1)!(k-1)!} \int_0^1 \frac{dt}{t} t^k \left(1 +  \frac{\bar{u}}{u}t \right)^{-1}  \nonumber \\
    & =-\frac{x^2M^2|b|^2}{4} \int_0^1 dt \frac{\bar{u}}{u+\bar{u}t} J_0\left( 2\rho\sqrt{t} \right) \, .
\end{align}
At the end, for those distributions containing  hypergeometric functions, 
the infinite series is replaced by an integral of a Bessel function of the first kind multiplied by a sufficiently regular function across the domain of integration.
The final result for the summed mass series of the helicity TMD~\eqref{eq:helicity} is 
\begin{equation}
    \frac{x^2M^2|b|^2}{2}\int_0^1 dt \frac{\bar{u}}{u+\bar{u}t} J_0\left( 2\rho\sqrt{t} \right)  - \rho\frac{u}{\bar{u}}J_1\left(2\rho\right)  \, .
\end{equation}

\section{Example of one-loop computation}
\label{app:one-loop_example}

In this appendix we present  a detailed example of a one-loop computation. 
To highlight the similarities and discrepancies with the standard parton-in-parton approach, this derivation may be compared with the results in Appendix B of Ref.~\cite{Bertone:2022frx}.

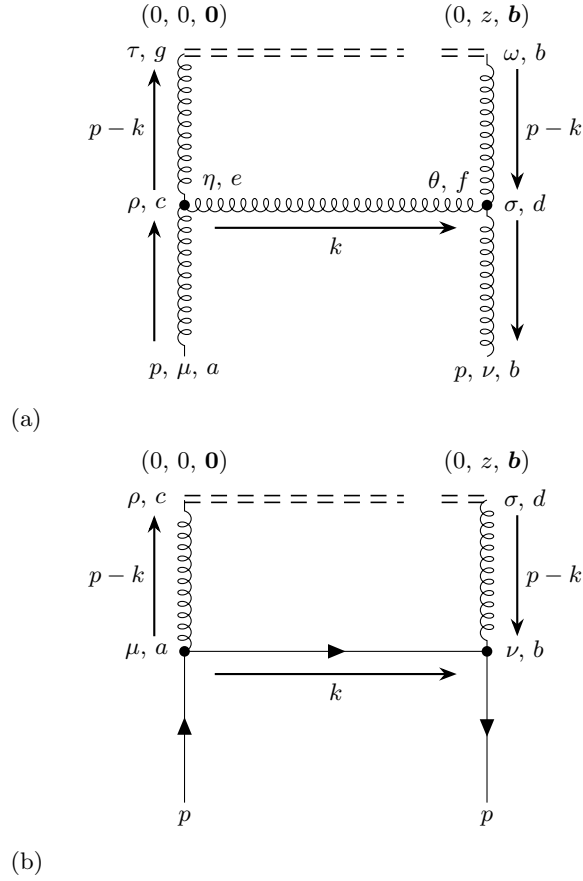
\begin{figure*}[htbp]
\centering
 \begin{subfigure}{0.5\textwidth}
        \centering
        \begin{tikzpicture}
            \begin{feynman}
                \vertex (i1) at (0,0);   
                \vertex (i2) at (0,2);   
                \vertex (a) at (0,4);   
                \vertex (b) at (4,4);   
                \vertex (i3) at (4,2);
                \vertex (f1) at (4,0);  
                \vertex (fr1) at (-0.4, 0.2);
                \vertex (fr2) at (-0.4, 1.8);
                \vertex (fr3) at (-0.4, 2.2);
                \vertex (fr4) at (-0.4, 3.8);
                \vertex (fr5) at (4.4, 3.8);
                \vertex (fr6) at (4.4, 2.2);
                \vertex (fr7) at (4.4, 1.8);
                \vertex (fr8) at (4.4, 0.2);
                \vertex (fr9) at (0.4, 1.7);
                \vertex (fr10) at (3.6, 1.7);

                \fill (i2) circle (2pt);
                \node at (0, 4.5) {$(0, \, 0, \, \boldsymbol{0})$};
                \node at (4, 4.5) {$(0,\, z,\, \boldsymbol{b})$};
                \node at (-0.5, 2) {$\rho, \, c$};
                \node at (0.5, 2.3) {$\eta, \, e$};
                \node at (3.5, 2.3) {$\theta, \, f$};
                \node at (4.5, 2) {$\sigma, \, d$};
                \node at (4, -0.2) {$p,\, \nu, \, b$};
                \node at (0, -0.2) {$p, \, \mu, \, a$};
                \node at (-0.5, 4) {$\tau, \, g$};
                \node at (4.5, 4) {$\omega, \, b$};
                \fill (i3) circle (2pt);
                \draw[->, thick, >=Stealth] (fr1) -- (fr2);
                \draw[->, thick, >=Stealth] (fr3) -- (fr4) node[midway, left] {$p-k$};
                \draw[->, thick, >=Stealth] (fr5) -- (fr6) node[midway, right] {$p-k$};
                \draw[->, thick, >=Stealth] (fr7) -- (fr8);
                \draw[->, thick, >=Stealth] (fr9) -- (fr10) node[midway, below] {$k$};

                \diagram*{
                    (i2) -- [gluon] (i1),
                    (a) -- [gluon] (i2),
                    (i2) -- [gluon] (i3),
                    (i3) -- [gluon] (b),
                    (f1) -- [gluon] (i3),
                };

                \draw [dash pattern=on 6pt off 4pt, semithick] (0, 4.06) -- (2.9, 4.06);
                \draw [dash pattern=on 6pt off 4pt, semithick] (0, 3.97) -- (2.9, 3.97);

                \draw [dash pattern=on 6pt off 4pt, semithick] (3.4, 4.06) -- (4.0, 4.06);
                \draw [dash pattern=on 6pt off 4pt, semithick] (3.4, 3.97) -- (4.0, 3.97);
            \end{feynman}
        \end{tikzpicture}
        \caption{}
        \label{fig:gluon}
  \end{subfigure}
    
    
  \begin{subfigure}{0.5\textwidth}
        \centering
        \begin{tikzpicture}
            \begin{feynman}
                \vertex (i1) at (0,0);   
                \vertex (i2) at (0,2);   
                \vertex (a) at (0,4);   
                \vertex (b) at (4,4);  
                \vertex (c) at (2.7,4);
                \vertex (d) at (4,2);
                \vertex (f1) at (4,0);
                \vertex (fr3) at (-0.4, 2.2);
                \vertex (fr4) at (-0.4, 3.8);
                \vertex (fr5) at (4.4, 3.8);
                \vertex (fr6) at (4.4, 2.2);
                \vertex (fr9) at (0.4, 1.7);
                \vertex (fr10) at (3.6, 1.7);

                \fill (i2) circle (2pt);
                \fill (d) circle (2pt);
                \node at (0, 4.5) {$(0, \, 0, \, \boldsymbol{0})$};
                \node at (4, 4.5) {$(0,\, z,\, \boldsymbol{b})$};
                \node at (4, -0.2) {$p$};
                \node at (0, -0.2) {$p$};
                \node at (-0.5, 2) {$\mu, \, a$};
                \node at (4.5, 2) {$\nu, \, b$};
                \node at (-0.5, 4) {$\rho, \, c$};
                \node at (4.5, 4) {$\sigma, \, d$};
                \draw[->, thick, >=Stealth] (fr3) -- (fr4) node[midway, left] {$p-k$};
                \draw[->, thick, >=Stealth] (fr5) -- (fr6) node[midway, right] {$p-k$};
                \draw[->, thick, >=Stealth] (fr9) -- (fr10) node[midway, below] {$k$};

                \diagram*{
                    (i1) -- [fermion] (i2),
                    (i2) -- [gluon] (a),
                    (i2) -- [fermion] (d),
                    (d) -- [fermion] (f1),
                    (b) -- [gluon] (d)
                };
                
                \draw [dash pattern=on 6pt off 4pt, semithick] (0, 4.06) -- (2.9, 4.06);
                \draw [dash pattern=on 6pt off 4pt, semithick] (0, 3.97) -- (2.9, 3.97);

                \draw [dash pattern=on 6pt off 4pt, semithick] (3.4, 4.06) -- (4.0, 4.06);
                \draw [dash pattern=on 6pt off 4pt, semithick] (3.4, 3.97) -- (4.0, 3.97);
            \end{feynman}
        \end{tikzpicture}
        \caption{}
        \label{fig:gluon-quark}
    \end{subfigure}
    \caption{One-loop diagrams for the residual parts of the gluon TMD matching relations.   Dashed double lines denote the Wilson lines. Diagram~\ref{fig:gluon} shows the contribution evaluated in Eq.~\eqref{eq:gluon-in-gluon-lg}, while diagram~\ref{fig:gluon-quark} refers to Eq.~\eqref{eq:gluon-in-quark-lg}.}
    \label{fig:one_loop_diagrams}
\end{figure*}

The parton-in-parton distributions in  Eqs.~\eqref{eq:gluon-in-gluon-lg} and~\eqref{eq:gluon-in-quark-lg} at one-loop correspond to the diagrams in Fig.~\ref{fig:one_loop_diagrams}. 

Using standard QCD Feynman rules, the diagrams evaluate to
\begin{align}
\label{eq:step1-gg}
    G_{gg}^{\Lambda,\mu\nu}(u) =& -\frac{xp_+}{2(N_c^2-1)} \int \frac{dz}{2\pi} e^{i\bar{u}zp_+} e^{i(bp_T)} R_{gg}^{\Lambda,\mu\nu}(z)\, , \\
\label{eq:step1-gq}
    G_{gq}^{\Lambda/\Gamma}(u) =& -\frac{xp_+}{2} \int \frac{dz}{2\pi} e^{i\bar{u}zp_+} e^{i(bp_T)} {\rm Tr} \left[ R_{gq}^\Lambda(z) \Gamma \right]\, ,
\end{align}
where
\begin{align}
    R_{gg}^{\Lambda,\mu\nu}(z) =& -4ia_{\rm s} f^{ace}f^{eca} \int \frac{d^{4-2\epsilon}k\mu^{2\epsilon}}{(2\pi)^{2-2\epsilon}} e^{-ik_+z}e^{i\boldsymbol{b}\cdot\boldsymbol{k_T}} \nonumber\\
    &\times V_{\mu\rho\eta}(p,k-p,-k)\, V_{\theta\sigma\nu}(k,p-k,-p) \nonumber\\
    &\times \frac{\Lambda^{\tau\omega}d^{\rho\tau}(p-k)d^{\eta\theta}(k)d^{\omega\sigma}(p-k)}{(k^2+i\epsilon)\left[ (p-k)^2+i\epsilon \right]^2} \, ,
\\
    R_{gq}^{\Lambda}(z) =& 4ia_{\rm s} C_F \int \frac{d^{4-2\epsilon}k\mu^{2\epsilon}}{(2\pi)^{2-2\epsilon}} e^{-ik_+z}e^{i\boldsymbol{b}\cdot\boldsymbol{k_T}} \nonumber\\
    & \times \frac{\gamma_\mu \slashed{k}\gamma_\nu d^{\mu\rho}(p-k)d^{\sigma\nu}(p-k)\Gamma_{\rho\sigma}}{(k^2+i\epsilon)\left[ (p-k)^2+i\epsilon \right]^2}\, ,
\end{align}
with
\begin{equation}
    V^{\mu\nu\rho}(q,l,r)=- g^{\mu\nu}(q-l)^\rho - g^{\nu\rho}(l-r)^\mu - g^{\rho\mu}(r-q)^\nu \, .
\end{equation}
The global phases in Eqs.~\eqref{eq:step1-gg} and~\eqref{eq:step1-gq} originate from (see also Ref.~\cite{Bertone:2022frx})
\begin{align}
    A_a^\mu(x)|k,s\rangle_g =& e^{-i(kx)}\epsilon_a^\mu(k,s)|0\rangle \, , \\
    \psi(x)|k,s\rangle_q =& e^{-i(kx)}u(k,s)|0\rangle \, .
\end{align}
Finally, for the gluon-in-quark case, the projectors onto the quark polarizations are $\Gamma\in \{ \slashed{p} \, , \gamma^5\slashed{p} \, , i\sigma^{(R+L)p}\gamma^5/2 \}$, which correspond, in order, to unpolarized, longitudinally polarized and transversely polarized quarks.

We will explicitly compute $G_{gg}^{U,\mu\nu}$. After integration over $k_+$ in Eq.~\eqref{eq:step1-gg},  we obtain
\begin{widetext}
\begin{align}
    G_{gg}^{U,\mu\nu} (u) =& 4ia_{\rm s} C_A \int \frac{d^{2-2\epsilon}\boldsymbol{k_T}\mu^{2\epsilon}}{(2\pi)^{2-2\epsilon}} \int \frac{dk^-e^{i\boldsymbol{b}\cdot\boldsymbol{k_T}}e^{i(bp_T)}}{(k^--k_1^-)(k^--k^-_2)^2} \Bigg\{ 2u(1-2u^2)(1-\epsilon)p_+^3 \bar{n}^\mu \bar{n}^\nu \nonumber \\ 
    &+ 8u(1-\epsilon)p_+ k_T^\mu k_T^\nu -2(3u-2+\epsilon u)p_+ p_T^\mu p_T^\nu  +2(1-2u)(\bar{u}+\epsilon u) p_+^2 \left( \bar{n}^\mu p_T^\nu + p_T^\mu \bar{n}^\nu \right) \nonumber \\
    &- \frac{2}{\bar{u}}\left(1-2u^2-2u\bar{u}\epsilon\right) p_+ \left( k_T^\mu p_T^\nu + p_T^\mu k_T^\nu \right) -\frac{2}{\bar{u}}(1-2u)(1-2u+2u^2-2u\bar{u}\epsilon)p_+^2\left( k_T^\mu n^\nu + n^\mu k_T^\nu \right) \nonumber \\
    & -\left[ 2(1+u^2)\frac{u}{\bar{u}}k^-p_+ -8\left( \frac{\bar{u}}{u} + \frac{u}{\bar{u}} \right)\boldsymbol{k_T}\cdot\boldsymbol{p_T} +\frac{u^3+3u^2-4u+4}{u\bar{u}}\boldsymbol{k}_{\boldsymbol{T}}^2 +4\bar{u}\left( \frac{\bar{u}}{u} + \frac{u}{\bar{u}} \right)\boldsymbol{p}_{\boldsymbol{T}}^2 \right] p_+g^{\mu\nu} \Bigg\} \, ,
\end{align}
\end{widetext}
where
\begin{align*}
    k_1^- =& \frac{\boldsymbol{k}_{\boldsymbol{T}}^2}{2\bar{u}P^+} -i\frac{\epsilon}{\bar{u}} \, , \\
    k_2^-=&\frac{2\boldsymbol{k_T}\cdot\boldsymbol{p_T}-\boldsymbol{k}_{\boldsymbol{T}}^2-\bar{u}\boldsymbol{p}_{\boldsymbol{T}}^2}{2uP^+} +i\frac{\epsilon}{u} \, .
\end{align*}
The integration over $k^-$ is evaluated using
\begin{align}
    \int \frac{dk^-}{(k^--k^-_1)(k^--k^-_{2})^2} =& 8\pi i \frac{u^2\bar{u}^2p_+^2}{\left( \boldsymbol{k}_{\boldsymbol{T}} - \bar{u}\boldsymbol{p_T} \right)^4} \, , \\
    \int \frac{dk^-k^-}{(k^--k^-_1)(k^--k^-_{2})^2} =& -4\pi i \frac{u^2\bar{u}p_+\boldsymbol{k}_{\boldsymbol{T}}^2}{\left( \boldsymbol{k}_{\boldsymbol{T}} - \bar{u}\boldsymbol{P_T} \right)^4} \, .
\end{align}
After integrating over  $k^-$ and performing the shift $\boldsymbol{k_T}\to \boldsymbol{k_T}+\bar{u}\boldsymbol{p_T}$, we obtain
\begin{widetext}
\begin{align}
    G_{gg}^{U,\mu\nu}(u) =& -8\pi a_{\rm s}C_A e^{iu(bp_T)} \int \frac{d^{2-2\epsilon}\boldsymbol{k_T}\mu^{2\epsilon}}{(2\pi)^{2-2\epsilon}} \frac{e^{i\boldsymbol{b}\cdot\boldsymbol{k_T}}}{\boldsymbol{k}_{\boldsymbol{T}}^4} \Bigg\{ 2\left( \frac{\bar{u}}{u} + \frac{u}{\bar{u}} \right)g^{\mu\nu} \nonumber \\
    & +(1-2u)(1-2u\bar{u}+2u\bar{u}\epsilon) \left[ p_+ \left( \boldsymbol{k}_{\boldsymbol{T}}^\mu \bar{n}^\nu + \bar{n}^\mu \boldsymbol{k}_{\boldsymbol{T}}^\nu \right) + \left( \boldsymbol{k}_{\boldsymbol{T}}^\mu \boldsymbol{p}_{\boldsymbol{T}}^\nu + \boldsymbol{p}_{\boldsymbol{T}}^\mu \boldsymbol{k}_{\boldsymbol{T}}^\nu \right) \right] \nonumber \\
    & -u\bar{u}(1-\epsilon) \left[ 4 \boldsymbol{k}_{\boldsymbol{T}}^\mu \boldsymbol{k}_{\boldsymbol{T}}^\nu +(1-2u)^2 \left( p_+^2 \bar{n}^\mu \bar{n}^\nu + p_+\left( \boldsymbol{p}_{\boldsymbol{T}}^\mu \bar{n}^\nu + \bar{n}^\mu \boldsymbol{p}_{\boldsymbol{T}}^\nu \right) \right) + \boldsymbol{p}_{\boldsymbol{T}}^\mu \boldsymbol{p}_{\boldsymbol{T}}^\nu \right] \Bigg\} \, .\label{G-gg_U}
\end{align}
\end{widetext}
Finally, we evaluate the integration over $\boldsymbol{k_T}$ using the following integrals 
\begin{align}
    P(\boldsymbol{b}^2) =& \int \frac{d^{2-2\epsilon}\boldsymbol{k_T}}{(2\pi)^{2-2\epsilon}} \frac{e^{i\boldsymbol{b}\cdot\boldsymbol{k_T}}}{\boldsymbol{k}_{\boldsymbol{T}}^2} = \frac{\left(\pi \boldsymbol{b}^2\right)^\epsilon}{4\pi} \Gamma(-\epsilon) \, , \label{integral-P}\\
    Q(\boldsymbol{b}^2) =& \int \frac{d^{2-2\epsilon}\boldsymbol{k_T}}{(2\pi)^{2-2\epsilon}} \frac{e^{i\boldsymbol{b}\cdot\boldsymbol{k_T}}}{\boldsymbol{k}_{\boldsymbol{T}}^4} = - \frac{\boldsymbol{b}^2\left(\pi \boldsymbol{b}^2\right)^\epsilon}{16\pi(1+\epsilon)} \Gamma(-\epsilon) \, ,\label{integral-Q}\\
    R^\mu (\boldsymbol{b}) =& \int \frac{d^{2-2\epsilon}\boldsymbol{k_T}}{(2\pi)^{2-2\epsilon}} \frac{e^{i\boldsymbol{b}\cdot\boldsymbol{k_T}}\boldsymbol{k}_{\boldsymbol{T}}^\mu}{\boldsymbol{k}_{\boldsymbol{T}}^4} = \frac{i\left(\pi \boldsymbol{b}^2\right)^\epsilon}{8\pi} \Gamma(-\epsilon) \boldsymbol{b}^\mu \, ,\label{integral-R}\\
    S^{\mu\nu} (\boldsymbol{b}) =& \int \frac{d^{2-2\epsilon}\boldsymbol{k_T}}{(2\pi)^{2-2\epsilon}} \frac{e^{i\boldsymbol{b}\cdot\boldsymbol{k_T}}\boldsymbol{k}_{\boldsymbol{T}}^\mu \boldsymbol{k}_{\boldsymbol{T}}^\nu}{\boldsymbol{k}_{\boldsymbol{T}}^4} \nonumber \\
    &= - \frac{\left(\pi \boldsymbol{b}^2\right)^\epsilon}{4\pi} \Gamma(-\epsilon) \left( \frac{g^{\mu\nu}_T}{2} - \frac{\boldsymbol{b}^\mu \boldsymbol{b}^\nu}{\boldsymbol{b}^2}\epsilon \right)\, .\label{integral-S}
\end{align}
The integral $Q$ in Eq.~\eqref{integral-Q} is evaluated using the $\alpha$-representation (see Ch.~2 of Ref.~\cite{Smirnov:2004ym}), while the integrals\eqref{integral-P},~\eqref{integral-R}, and~\eqref{integral-S} are obtained by differentiating $Q$ with respect to $\boldsymbol{b}$. 
To simplify the expression of the final result, we have introduced 
in Eq.~\eqref{G-gg_U} appropriate terms proportional to $n^\mu$ and $n^\nu$. Since these terms are orthogonal to the projectors defined in Eqs.~\eqref{eq:unpolarized-gluon-projector}-\eqref{eq:linearly-polarized-gluon-projector},  they do not affect the extraction of the matching coefficients. 
The final result for $G_{gg}^{U,\mu\nu}$, at the end of these operations, reads 
\begin{align}
    G_{gg}^{U, \mu\nu} (u,b) =& C_{gg} \Bigg\{ 4\frac{(1-u\bar{u})^2}{u\bar{u}} g^{\mu\nu}_T - 8u\bar{u}\epsilon \left( \frac{g^{\mu\nu}_T}{2} - \frac{b^\mu b^\nu}{b^2} \right) \nonumber \\
    & + i(1-2u)\left( 1 - 2u\bar{u} +2u\bar{u}\epsilon \right) \left[ b^\mu p^\nu + p^\mu b^\nu  \right] \nonumber \\
\label{eq:unpolarized-gluon-in-gluon}
    & - \frac{b^2}{2}u\bar{u}(1-2u)^2(1-2\epsilon) p^\mu p^\nu \Bigg\} e^{iu(bp_T)} \, .  
\end{align}
where
\begin{equation}
    C_{gg}=a_{\rm s}C_A \left(\pi \mu^2 \boldsymbol{b}^2 \right)^\epsilon \Gamma(-\epsilon) \, .
\end{equation}
By contracting with $d^{\mu\nu}(p)$, $-i\epsilon^{\mu\nu}(p)$ and $t^{\mu\nu}(p)$ given in Eqs.~\eqref{eq:unpolarized-gluon-projector}-\eqref{eq:linearly-polarized-gluon-projector}, we can extract the splitting function $p_{gg}$ along with  the residual functions $R_{gg}^{U/U}$, $R_{gg}^{U/L}$ and $R_{gg}^{U/T}$. 

The transverse case is particularly interesting, as it explicitly demonstrates the cancellation of the undesired $b^2p_T^2$ terms in the final expression of the matching coefficients. In fact, by evaluating the contraction with the leading, suppressed, and doubly suppressed components of Eq.~\eqref{eq:linearly-polarized-gluon-projector} separately, we obtain
\begin{widetext}
\begin{align}
    \frac{G_{gg}^{U, \mu\nu} (u,b)}{2C_{gg}}\left( R^\mu R^\nu + L^\mu L^\nu \right) =& \Bigg[ -4\frac{(bL)^2+(bR)^2}{b^2} u\bar{u}\epsilon -i\left( (bL)(p_TL) + (bR)(p_TR) \right) (1-2u)(1-2u\bar{u}+2u\bar{u}\epsilon)\nonumber \\
    & + \frac{b^2}{4} \left( (p_TL)^2 + (p_TR)^2 \right) u\bar{u}(1-2u)^2(1-2\epsilon) \Bigg] e^{iu(bp_T)} \, ,\\
    \frac{G_{gg}^{U, \mu\nu} (u,b)}{2C_{gg}}\frac{\tilde{p}_{T,\mu}n_\nu + n_\mu \tilde{p}_{T,\nu}}{p_+} =& \Bigg[ i\left( (bL)(p_TL) + (bR)(p_TR) \right) (1-2u)(1-2u\bar{u}+2u\bar{u}\epsilon) \nonumber \\
    & - \frac{b^2}{2} \left( (p_TL)^2 + (p_TR)^2 \right) u\bar{u}(1-2u)^2(1-2\epsilon) \Bigg] e^{iu(bp_T)}\, , \\
    \frac{G_{gg}^{U, \mu\nu} (u,b)}{2C_{gg}}\frac{(p_TL)^2+(p_TR)^2}{p_+^2} n_\mu n_\nu =& \frac{b^2}{4} \left( (p_TL)^2 + (p_TR)^2 \right) u\bar{u}(1-2u)^2(1-2\epsilon) e^{iu(bp_T)} \, .
\end{align}
\end{widetext}
Only by summing these three contributions (that is, by performing the contraction with the complete expression for $t^{\mu\nu}(p)$), are the undesired $b^2p_T^2$ terms removed, leading to the final result
\begin{equation}
    R_{gg}^{U/T} (u) = -4u\bar{u} C_A e^{iu(bp_T)},
\end{equation}
which is in full agreement with the result found in Ref.~\cite{Bertone:2025vgy}.
Such an exact cancellation occurs for all the residual functions $R_{ij}^{\Lambda/\Gamma}$. Furthermore, this example explicitly demonstrates that the matching coefficients in the extended parton-in-parton framework are related to the standard ones by a simple phase factor:  $\mathcal{C}_{ext}=\mathcal{C}_{std}e^{iu(bp_T)}$. We argue that a similar ansatz may hold  at higher perturbative orders.

\bibliography{biblio}

\end{document}